\documentclass[12pt]{amsart}
\usepackage[utf8]{inputenc}

\usepackage{mathptmx}      
\usepackage{fullpage, bbm, enumerate}
\usepackage{latexsym}
\usepackage[pdftex]{graphicx}
\usepackage{epstopdf}
\usepackage{latexsym,amsmath,amsthm,amsfonts,amscd,graphicx,pstricks,pst-plot}
\usepackage{indentfirst}
\usepackage{epsfig}
\usepackage{amssymb}
\usepackage{setspace}
\usepackage{mathrsfs}
\usepackage{caption}
\usepackage{subcaption}

\newtheorem{ex}{Example}[section]
\newtheorem{theorem}{Theorem}[section]
\newtheorem{lemma}[theorem]{Lemma}
\newtheorem{remark}[theorem]{Remark}
\newtheorem{ass}[theorem]{Assumption}

\newcommand{\ignore}[1]{}

\newcommand{\EE}{\mathbb{E} }

\newcommand{\PP}{\mathbb{P} }
\newcommand{\QQ}{\mathbb{Q} }
\newcommand{\RR}{\mathbb{R} }

\begin{document}

\title{Polynomial term structure models}
\author{Si Cheng \and Michael R. Tehranchi}

\maketitle

\begin{abstract}
We explore a class of tractable interest rate models that have the property
that the prices of zero-coupon bonds can be expressed as polynomials of a state diffusion
process.  These models are arbitrage free in the sense that prices of zero-coupon bonds with all maturities are simultaneously local martingales
under the risk neutral measure.
Our main result is a classification of such models in the spirit of Filipovic's
maximal degree theorem  for exponential polynomial models. In particular for the scalar factor models, we also characterise such models where bonds prices are true martingales. Due to the fact that the bond prices are tractable, these models are easy to calibrate in general.
\end{abstract}

\section{Introduction}
A factor model of the interest rate term
structure is one in which the time-$t$ spot interest rate is of the form
$$
r_t = R(Z_t)
$$
and the time-$t$ price of a bond of maturity $T$ is of the form
$$
P_t(T) = H(T-t, Z_t)
$$
where $R:\RR^d \to \RR$ and $H:\RR^+ \times \RR^d \to \RR$ are given functions and $Z = (Z_t)_{t \ge 0}$
is a $d$-dimensional factor process.  Here we consider only bonds which pay no coupons, suffer
no default risk, and have unit face value.
To match the terminal price, we assume that
\begin{equation}\label{eq:bc}
H(0, z) = 1 \mbox{ for all } z.
\end{equation}
More importantly, to ensure that there is no arbitrage, one assumes the existence
of a probability measure  $\QQ$ under which the discounted bond prices, defined by
$$
\tilde P_t(T)   = e^{-\int_0^t r_s ds} P_t(T),
$$
are local martingales.
Of course, this assumption imposes a constraint on the functions $R$ and $H$ and the dynamics of $Z$ under
$\QQ$.   Indeed, in the case $d = 1$, if $Z$ is assumed to be a
solution of the stochastic differential equation
\begin{equation}\label{eq:SDE}
dZ_t =  b(Z_t) dt + \sigma(Z_t) dW_t,
\end{equation}
where $W$ is a scalar Brownian motion and $b$ and $\sigma$ are given functions,
then It\^o's formula yields the appropriate consistency condition
\begin{equation}\label{eq:PDE}
\partial_x H = b \ \partial_z H + \frac{1}{2} \sigma^2 \ \partial_{zz} H - R H \mbox{ for } x > 0, z \in I
\end{equation}
where $I \subseteq \RR$ is the state space of the process $Z$.
In principle, the above partial differential equation \eqref{eq:PDE}
with boundary condition \eqref{eq:bc}
 can be solved numerically  whenever the functions
$b$, $\sigma$ and $R$ are suitably well-behaved.  However, to actually implement such a model, one
must first calibrate the parameters, and unfortunately, resorting to a numerical methods at
this stage can   obscure the relationship between the dynamics of the factor process and the resulting bond prices.
Therefore, there has been considerable
interest in developing tractable models, where the function $H$ is of reasonably explicit
form.

Perhaps the two most famous tractable factor models are those of Vasicek \cite{V} and Cox, Ingersoll \& Ross \cite{CIR}.
In these models the factor process is identified with the spot interest rate, so in the notation above, $R(z)=z$,
the functions $b$ and $\sigma^2$ are assumed to be affine, and the function $H$ is of the exponential
affine form
$$
H(x,z) = e^{h_0(x)  + h_1(x) z }.
$$
It is easy to see that the consistency equation \eqref{eq:PDE} reduces to a system of coupled Riccati
ordinary differential equations for the functions $h_0$ and $h_1$ with boundary conditions $h_0(0)=h_1(0) = 0$.
 Duffie \& Kan \cite{DK} studied exponential affine models where the factor process is of arbitrary dimension $d \ge 1$,
leading to much study of the properties of these models by a number of researchers.  A notable contribution to this literature
is the general characterisation of exponential affine term
structure models by Duffie, Filipovic \& Schachermayer \cite{DFS}.

An exponential affine model can be considered a special case of the family of exponential
quadratic models.  An early example of an exponential quadratic model was proposed by Longstaff \cite{L},
and has since been developed and generalised by Jamshidian \cite{J}, Leippold \&  Wu \cite{LW},
and  Chen, Filipovic \& Poor \cite{CFP} among others.

One may wonder if there exist non-trivial
 exponential polynomial models of arbitrary degree. Filipovic answered this
question in the negative, by showing that the maximal degree for exponential polynomial models
is necessarily two.  That is to say, the exponential quadratic models are indeed
the most general class of exponential polynomial models.

Formally speaking, one may consider the function $H$ of the exponential polynomial models as an infinite series of the powers of the second argument $z^k$ by using Taylor's expansion. In this chapter, we generalise the exponential polynomial models by considering the class of polynomial models, of which the function $H(x,z)$ will be a finite sum of powers of the second argument $z$. In the case where the factor process is scalar-valued, the function $H$ is of the form
\begin{equation}\label{eq:poly}
H(x, z) = \sum_{k=0}^n g_k(x) z^k
\end{equation}
for $n+1$ differentiable functions $g_k: \RR^+ \to \RR$.  The main result is a classification of all such
models when the factor process is assumed to satisfy an SDE of the form of
equation \eqref{eq:SDE}.  It turns out that the functions $b$, $\sigma$ and $R$
are necessarily polynomials of low degree and the functions $g_k$ solve a system
of coupled linear ODEs.  In light of Filipovic's maximal degree theorem for exponential polynomial models, it might come as a
surprise the degree $n$ is not constrained;
indeed, the exponential quadratic models can be seen as the $n \to \infty$ limit, in a certain sense, of the polynomial models.

This work is inspired by the interest rate model of Siegel \cite{S}.  He
showed that for all integers $d \ge 1$ there exists explicit functions $b:\RR^d \to \RR^d$,
$\sigma: \RR^d \to \RR^{d\times d}$, $R: \RR^d \to \RR$ and $H:[0, \infty) \times \RR^d \to \RR$,
depending on $d$ parameters such that if $Z$ is a solution of the stochastic differential equation
$$
dZ_t= b(Z_t) dt + \sigma(Z_t) dW_t
$$
where $W$ is a $d$-dimensional Brownian motion,
$$
r_t = R(Z_t)
$$
and
$$
P_t(T) = H(T-t, Z_t),
$$
then the processes $\tilde P(T)$ are martingales for each $T \ge 0$, where
$$
\tilde P_t(T) = e^{-\int_0^t r_s ds} P_t(T).
$$
Furthermore, the functions $a:=\sigma \sigma^\top$ and $b$ are quadratic and the functions $R$ and $H(x, \cdot)$
are affine for all $x \ge 0$, and $H(0, z) = 1$ for all $z \in \RR^d$.
  In particular, the random variables $P_t(T)$ constitutes an arbitrage-free bond
	price model, where $r_t$ is the corresponding spot interest rate.

A related work is that of Cuchiero, Keller-Ressel \& Teichmann \cite{CKT},
who characterise a class of time-homogeneous Markov process $Y$  with the property that the $n$-th (mixed)
 moments can be expressed as a polynomial of initial point $Y_0$ of degree at most $n$.
Indeed, consider the $d=1$ case and let $F_n$ be the family of polynomials of degree at most $n$:
\begin{equation}\label{eq:F}
F_n = \left\{f: f(z) = \sum_{k=0}^n f_k z^k, \  f_k \in \RR \right\}.
\end{equation}
They study the processes $Y$ that have the property that
for all $t \ge 0$ and for \textit{any} degree $n$ and \textit{any} polynomial $g \in F_n$,  there exists
a polynomial $h \in F_n$ such that
\begin{align*}
\mathbb{E}[g(Y_t)|Y_0 = y] = h(y).
\end{align*}
In contrast, in this work we study  processes $Z$ such that for all $t \ge 0$ there
exists a polynomial $h = H(t,\cdot) \in F_n$ such that
$$
\EE[ e^{-\int_0^t R(Z_s) ds} | Z_0 = z] = h(z)
$$
where the function $R$ and the degree $n$ are \textit{fixed}.
For example, the example (iii) on page 721 of Cuchiero's \cite{CKT} paper. They show that the process
\begin{align*}
dZ_t = (\frac{1}{2}-bZ_t+\frac{1}{2}Z_t^2)dt + \sqrt{Z_t^2 (1-Z_t)}dW_t
\end{align*}
is not a 3-polynomial. However $Z$ may serve as the factor process in the polynomial model when $n=3$.
In particular, their results
do not imply ours, or vice versa. For further applications of polynomial preserving processes to finance, consult the recent paper of Filipovic and Larsson \cite{FL}.


The remainder of this paper is arranged as follows.  In section \ref{se:results},
we present the main result, a classification of interest rate models in which the
bond price can be expressed as a polynomial of a scalar factor process.  In section \ref{se:examples}, we
consider two concrete  examples of this class of models and
further analyse their properties. Finally in section \ref{se:extensions}, we briefly discuss two extensions: a Hull-White-type extension where the coefficients are allowed to be time dependent, and the case when $d > 1$.

\section{The main results} \label{se:results}
To more clearly see the structure of the argument we consider only the case $d=1$ in this section. The multi-dimensional case is considered in section 4. This section contains the main result of this paper, a classification of polynomial
term structure models.  We fix a degree $n \ge 1$, and
 let $H: \RR^+ \times \RR \to \RR$ be of a polynomial in the second
variable as in equation \eqref{eq:poly}. To match the boundary condition \eqref{eq:bc} we will assume
\begin{equation}\label{eq:bc-g}
g_0(0) = 1 \mbox{ and } \ g_k(0) = 0 \mbox{ for all } 1 \le k \le n.
\end{equation}

\begin{ass} \label{as:main}
We will assume that the coefficient functions
$(g_k)_k$ are differentiable and linearly independent.
We also assume that the scalar factor process $Z$ is a non-explosive solution
of the SDE \eqref{eq:SDE} with continuous drift and volatility function $b,\sigma$ where the state space $I \subseteq \RR$ is a bounded open
interval.
\end{ass}
For ease of notation, we will define
$$
a(z) := \sigma^2(z)
$$
and use $a(z)$ and $\sigma^2(z)$ interchangeably.
\begin{remark} \label{th:assump}
The above assumptions play an important role in the class of polynomial models. We will postpone the discussion of these assumptions to the end of this section.
\end{remark}

\begin{theorem}\label{th:main}  The function $H$ satisfies the PDE \eqref{eq:PDE}
if and only if the following conditions hold true:

\noindent Case $n=1$.
\begin{enumerate}[(A)]
\item   $ R(z) = R_0 + R_1 z$ and $b(z) = b_0 + b_1 z + b_2 z^2$
where $R_1= b_2$.
\item  $(g_0, g_1)$ is the unique solution to the system of linear ODEs
\begin{align*}
\dot g_0 &= - R_0 g_0 + b_0 g_1 \\
\dot g_1 &= -R_1 g_0 + (b_1-R_0) g_1
\end{align*}
subject to the boundary conditions \eqref{eq:bc-g}.
	\end{enumerate}
	
	\noindent Case $n \ge 2$.
\begin{enumerate}[(A)]
\item  $R(z) = R_0 + R_1 z + R_2 z^2$, $b(z) = b_0+ b_1 z + b_2 z^2 + b_3 z^3$ and
$\sigma^2(z) = a_0+ a_1 z + a_2 z^2 + a_3 z^3 + a_4 z^4$
where the coefficients are such that
$$
 R_2 =  \tfrac{n}{2} b_3 = - \tfrac{n(n-1)}{2} a_4 \mbox{ and } R_1 = n b_2 + \tfrac{n(n-1)}{2} a_3.
$$
\item  $(g_0, \ldots, g_n)$ is the unique solution to the system of linear ODEs
\begin{align*}
\dot{g_k} = & g_{k-2} \left( (k-2) b_3 + \frac{(k-2)(k-3)}{2} a_4 -R_2 \right) +
g_{k-1} \left( (k-1) b_2 + \frac{(k-1)(k-2)}{2} a_3 - R_1 \right) \\
& + g_k \left( kb_1 + \frac{k(k-1)}{2} a_2 - R_0 \right) +
  g_{k+1} \left( (k+1)b_0 + \frac{k(k+1)}{2} a_1 \right) +  g_{k+2} \frac{(k+2)(k+1)}{2} a_0
\end{align*}
subject to the boundary conditions \eqref{eq:bc-g}, where we interpret $g_{-2} = g_{-1} = g_{n+1} = g_{n+2} = 0$.
	\end{enumerate}

\end{theorem}

Before proceeding to the proof, we pause for several remarks.

\begin{remark}
The solution of the system of ODEs appearing in condition (B) of Theorem \ref{th:main} can be equivalently
described as follows.   Let  $S = (S_{i,j})_{i,j = 0}^n$ be the $(n+1) \times (n+1)$ matrix with entries
$$
S_{j+k,j} = j b_{k+1} + \tfrac{j(j-1)}{2}  a_{k+2} - R_{k}
$$
and where $R_k= b_k = a_k= 0$ when $k < 0$ and $R_k = b_{k+1} = a_{k+2} = 0$ when $k > 2$.  For instance,
when $n \ge 4$, the matrix has the form
$$
S = \left( \begin{array}{clllll}
-R_0 & \quad b_0          & \quad \hskip .85cm  a_0    &                        &           				  &   \\
-R_1 & \quad b_1- R_0     & \quad 2b_0 + a_1          & \quad \hskip .85cm  3 a_0     &           				  &   \\
-R_2 & \quad b_2 -R_1     &  \quad 2b_1+ a_2 -R_0     & \quad 3b_0 + 3a_1            & \quad \hskip .85cm  6 a_0 &   \\
     & \quad b_3 -R_2     & \quad 2b_2 + a_3 - R_1    & \quad 3b_1+3a_2 -R_0         & \quad 4b_0 + 6 a_1 			& \ddots \\
     &              & \quad 2 b_3 + a_4  -R_2   & \quad 3b_2 + 3 a_3 -R_1      & \quad 4 b_1 + 6 a_2 -R_0 & \ddots \\
     &              &                     & \quad \hskip .9cm \ddots     & \quad \hskip .9cm \ddots & \ddots
		\end{array} \right).
$$
Now letting
$$
G(x) = \left( \begin{array}{c} g_0(x) \\ g_1(x) \\ \vdots \\ g_n(x) \end{array} \right).
$$
The ODE becomes
$$
\dot G = S G,
$$
and, in particular, the solution can be expressed as
$$
G(x) = e^{Sx} G(0),
$$
where the boundary condition is given by
$$
G(0) = \left( \begin{array}{c} 1 \\ 0 \\ \vdots \\ 0 \end{array} \right).
$$

\end{remark}

\begin{remark} \label{re:pfun}
Assuming that $S$ has $n+1$ distinct real eigenvalues $\lambda_0, \ldots, \lambda_n$,
 we know from elementary linear algebra that we can express $G$ via
$$
G(x) = \sum_{i=0}^n p_i e^{\lambda_i x}
$$
for a collection of $n+1$ vectors $(p_i)_i$ in $\RR^{n+1}$.   Hence the bond pricing function is
of the form
$$
H(x,z) = \sum_{i=0}^n P_i(z) e^{\lambda_i x}
$$
where the function
$$
P_i(z) = \sum_{k=0}^n p_{i,k} z^k
$$
is the polynomial whose coefficients are given by the vector $p_i$.
That is to say, the bond price can be seen to be a linear combination of the bond prices
arising from models with constant interest rates $r = -\lambda_i$, where the coefficients
of the combination depend on the factor process.   Also note that generically the
long maturity interest rate in this model is given by
$$
\lim_{x \to \infty} - \frac{1}{x} \log H(x,z) = - \max_i  \lambda_i.
$$
\end{remark}


\begin{remark}
Notice that the eigenvalues of the matrix $S$ are the zeros of the characteristic polynomial
which has  degree $n+1$.  It is well know that there exists an explicit formula, discovered by Ferrari in 1540,
 for the zeros of quartic polynomials, and hence
the eigenvalues of $S$ can  be expressed in a closed formula in terms of the matrix entries when $n \le 3$.
In particular, in this case, the functions $g_i$ can be written, at least in principle, in terms of the model parameters.

 When $n \ge 4$, there is little hope for explicit formulae for the functions $g_i$ in
terms of the model parameters.  However, note that the matrix is sparse, in the sense that there are
at most five non-zero matrix entries per row.  In particular, the product of the matrix exponential $e^{Sx}$
and the vector $G(0)$ can be computed efficiently, and hence the lack of explicit formulae is not necessarily
a prohibitive disadvantage.
\end{remark}

\begin{remark}  When $n \ge 2$, it seems as though there are ten free parameters: $n$, $R_0$, $b_0, \ldots, b_2$,
and $a_0, \ldots, a_4$.  However, since we are really interested in the interest rate, but not the
factor process, and since the function $R$ is quadratic, we need only consider two subclasses of models.

Indeed, if $R_2= 0$ so that the function $R$ is affine, we can make a change of variables so that
$R(z) =  z$, and hence  the
factor $Z$ can be identified with the short rate $r$.   Also note that $b_3=a_4 =0$ and hence the
 SDE for $r$ is of the
seven parameter 3/2-type model family (assuming solutions exist)
$$
dr_t = (b_0 + b_1 r_t + \tfrac{c}{n} r_t^2) dt + \sqrt{a_0 + a_1 r_t + a_2 r_t^2 + \tfrac{2(1-c)}{n(n-1)}  r_t^3} dW_t.
$$
Note that by setting $a_1 = a_2 = 0$ we see that this polynomial term structure model approximates,
in some sense, an exponential affine model when $n$ is large.
\ignore{
and the $(n+1)\times (n+1)$ matrix $S$ is
$$
S = \left( \begin{array}{clllll}
 0 & b_0          & \hskip .9cm  a_0    &                        &           				  &   \\
-1 & b_1      		& 2b_0 + a_1          & \hskip .9cm  3 a_0     &           				  &   \\
  & b_2 -1    		&  2b_1+ a_2       & 3b_0 + 3a_1            & \hskip .9cm  6 a_0 &   \\
     & \tfrac{c}{n}    & 2b_2 + \tfrac{2(1-c)}{n(n-1)} - 1    & 3b_1+3a_2           & 4b_0 + 6 a_1 			& \ddots \\
     &              &   \tfrac{2c}{n}        & 3b_2 +  \tfrac{6(1-c)}{n(n-1)} - 1      & 4 b_1 + 6 a_2   & \ddots \\
     &              &                     & \hskip .9cm \ddots     & \hskip .9cm \ddots & \ddots
		\end{array} \right),
$$
}

Otherwise, if $R_2 \ne 0$, we can make another change of variables so that $R(z) = R_0 \pm z^2$.  The factor process
then evolves according to one of the following seven parameter SDEs (assuming solutions exist)
$$
dZ_t = (b_0 + b_1 Z_t  \pm \tfrac{1}{n}Z_t^2 ( 2 Z_t - c) ) dt +  \sqrt{a_0 + a_1 Z_t +
 a_2 Z_t^2 \mp \tfrac{2}{n(n-1)} Z_t^3 ( Z_t- c) } dW_t.
$$

Note that in both cases, the parameter $n$ must be an integer greater than one.  In particular,
calibration of such models to financial data must impose this constraint.

Note also that the equations $\frac{c}{n} = b_2$ and $\frac{2(1-c)}{n(n-1)} = a_3$ does not uniquely identify the pair $(c,n)$. For instance, \newline $(c,n) = (\frac{7}{6}, 2)$ and $(c,n) = (\frac{7}{2}, 6)$ correspond to the same spot rate model:
$$
dr_t = \tfrac{7}{12}r_t(r_t - 2)dt + r_t \sqrt{\tfrac{1}{6}(1-r_t)}dW_t
$$
In particular, this model is of degree two not degree six, since the functions $g_3 = g_4 = g_5 = g_6 = 0$ are not linearly independent.
\end{remark}

\begin{remark}  In the case $n=1$, the affine function $R$ is monotone, so there is no loss of generality taking $R_0 = 0$ and $R_1 = 1$.
In this case $r_t = Z_t$ and the short rate model becomes
$$
r_t = (b_0 + 2c r_t + r_t^2) dt + \sigma(r_t) dW_t.
$$
In this case, the functions $g_0$ and $g_1$ can be computed explicitly:
$$
g_0(x) = [ \cosh (  q x)  - \tfrac{c}{q} \sinh( q x) ] e^{cx }
$$
and
 $$
g_1(x) =   - \frac{1}{q} \sinh(   qx)  e^{c  x }
$$
where
$$
q=  \sqrt{ c^2 - b_0}.
$$
The bond pricing function is then
$$
H(x,r) =  \frac{1}{2}[1 + (c-q) r] (1+ c/q) e^{ (c-q) x} + \frac{1}{2}[1 + (c+q) r] (1- c/q) e^{ (c+q) x}.
$$

\ignore{
$$
g_0(x) =  \frac{q_+ e^{   q_- x } - q_- e^{   q_+ x}} {\sqrt{ b_1^2 - 4b_0}}
$$
and
$$
g_1(x) = b_0 \frac{e^{   q_- x } - e^{   q_+ x}} {\sqrt{ b_1^2 - 4b_0}}
$$
where
$$
 q_{\pm} = \frac{ b_1 \pm  \sqrt{ b_1^2 - 4b_0}}{2}
$$
are the eigenvalues of the matrix
$$
S = \left( \begin{array}{cc}  0 & b_0 \\ - 1 & b_1 \end{array} \right),
$$
}
As noted above,
 this calculation is independent of the function $\sigma$ (as long as the SDE has non-explosive bounded solutions). That is to say, set of current bond prices is not sufficient to fully calibrate the model. The parameter $\sigma$ could, in principle be estimated from historical data. Alternatively it could be calibrated from other interest derivatives.
\ignore{
$$
P_t(t+x) = \frac{q_+ e^{q_-x}}{q_+-q_-}(1+q_- r_t) - \frac{q_- e^{q_+x}}{q_+-q_-}(1+q_+ r_t)
$$
where $q_{\pm}$  are the eigenvalues of the matrix
}
\end{remark}

We are now ready to present the proof of Theorem \ref{th:main}.
\begin{proof}
Let
$$
A_k(z) = k b(z)  z^{k-1} + \frac{k(k-1)}{2} \sigma^2(z) z^{k-2} - R(z) z^k.
$$
Equation \eqref{eq:PDE} holds if and only if the equation
\begin{equation}\label{eq:identity}
\sum_{k=0}^n \dot g_k(x) z^k  = \sum_{k=0}^n g_k(x) A_k(z)
\end{equation}
holds identically.

	\ignore{
For sufficiency, it is straightforward to verify that if $R$, $b$ and (when $n\ge 2$) $\sigma^2$ are polynomials with coefficients
satisfying the conditions (A) and that the functions
$( g_0(x), \ldots, g_n(x) )$ satisfies the system of differential
equation  defined by (B), then equation \eqref{eq:identity} holds.
}

We first show that if
equation \eqref{eq:identity} holds then the functions $A_k \in F_n$ for $k$, where $F_n$ are the polynomials
of degree at most $n$ defined in equation \eqref{eq:F}.
 To see this, use the assumed
linear independence of the functions $(g_i)_i$ to pick $n+1$ points
$0 \le x_0 < \ldots < x_n$ such that the $(n+1)\times (n+1)$ matrix
$ (g_i(x_j))_{i,j}$ is invertible.   By evaluating equation \eqref{eq:identity}
at the points $(x_j)_j$ we get the following matrix representation:
\begin{align*}
\left(
  \begin{array}{ccc}
    g_0(x_0) & \cdots & g_n(x_0) \\
    \vdots & \ddots & \vdots \\
    g_0(x_n) & \cdots & g_n(x_n) \\
  \end{array}
\right)
\left(
  \begin{array}{c}
    A_0(z) \\
    \vdots \\
    A_n(z) \\
  \end{array}
\right) =
\left(
  \begin{array}{c}
    \sum_{k=0}^n \dot{g}_k(x_0) z^k \\
    \vdots \\
    \sum_{k=0}^n \dot{g}_k(x_n) z^k \\
  \end{array}
\right)
\end{align*}
and solve for the $A_i(z)$, we see that $A_i(z)$ is
a linear combination of monomials $z^k$ of degree at most $n$.

\noindent \textit{Case} $n=1$.  Note that
\begin{align*}
R(z) & = -A_0(z) \\
b(z) & = A_1(z) + z R(z).
\end{align*}
Since $A_0$ and $A_1$ are in $F_1$, i.e. are affine,  then $R$ is affine and $b$ is quadratic.  Letting
$b(z) = b_0 + b_1 z + b_2 z^2$ and $R(z) = R_0 + R_1 z$ the above system equation
implies $b_2 = R_1$.  Finally, the identity \eqref{eq:identity} becomes
$$
\dot g_0 + \dot g_1 z = g_0 (  R_0 + R_1 z) + g_1 ( b_0 + (b_1 - z R_0)z ).
$$
Equating coefficients of $z$ yields the necessity and sufficiency of the system of ODEs.

\noindent \textit{Case} $n \ge 2$.  Note that
\begin{align*}
R(z) & = -A_0(z) \\
b(z) & = A_1(z) + z R(z) \\
\sigma^2(z) &= A_2(z) - 2 z b(z) + z^2 R(z).
\end{align*}
Since the functions $A_i$ are polynomials, so are the functions $R$, $b$, and $\sigma^2$.
On the other hand
\begin{align*}
A_n(z) & = n b(z)  z^{n-1} + \frac{n(n-1)}{2} \sigma^2(z) z^{n-2} - R(z) z^n \\
& = z^{n-2} \left( n b(z) z  + \frac{n(n-1)}{2} \sigma^2(z)  - R(z) z^2 \right) \in F_n
\end{align*}
and, since the term in brackets is a polynomial, we have
\begin{equation}\label{eq:n}
 n b(z) z  + \frac{n(n-1)}{2} \sigma^2(z)  - R(z) z^2 \in F_2 \subseteq F_4.
\end{equation}
Similarly, since $A_{n-1} \in F_n$ and $A_{n-2} \in F_n$ we have
\begin{align}
(n-1) b(z) z  + \frac{(n-1)(n-2)}{2} \sigma^2(z)  - R(z) z^2 & \in F_3 \subseteq F_4  \label{eq:n-1} \\
(n-2) b(z) z  + \frac{(n-2)(n-3)}{2} \sigma^2(z)  - R(z) z^2 & \in F_4.  \label{eq:n-2}
\end{align}
Since
\begin{align*}
\sigma^2(z) = & \left( n b(z) z  + \frac{n(n-1)}{2} \sigma^2(z)  - R(z) z^2 \right) +
\left( (n-2) b(z) z  + \frac{(n-2)(n-3)}{2} \sigma^2(z)  - R(z) z^2 \right)  \\
& -  2\left( (n-1) b(z) z  + \frac{(n-1)(n-2)}{2} \sigma^2(z)  - R(z) z^2 \right)
\end{align*}
inclusions \eqref{eq:n}, \eqref{eq:n-1} and \eqref{eq:n-2} together yield
\begin{equation}\label{eq:sigma}
\sigma^2 \in F_4
\end{equation}
Similarly, since
\begin{align*}
zb(z) = &  \left( n b(z) z  + \frac{n(n-1)}{2} \sigma^2(z)  - R(z) z^2 \right) -
\left( (n-1) b(z) z  + \frac{(n-1)(n-2)}{2} \sigma^2(z)  - R(z) z^2 \right) \\
& - (n-1) \sigma^2
\end{align*}
inclusions \eqref{eq:n}, \eqref{eq:n-1} and \eqref{eq:sigma} together yield
\begin{equation}\label{eq:b}
b \in F_3.
\end{equation}
Finally,  inclusions \eqref{eq:n}, \eqref{eq:sigma} and \eqref{eq:b} together yield
$$
R \in F_2.
$$

Recall that $A_n$ is of degree at most $n$.
Now substituting
 $R(z) = \sum_{k=0}^2 R_k z^k, \ b(z) = \sum_{k=0}^3 b_k z^k, \ \sigma^2(z) = \sum_{k=0}^4 a_k z^k$
into the definition of $A_n$, and setting the coefficient of $z^{n+2}$ to zero yields
\begin{equation}\label{eq:first}
  n b_3 + \frac{n(n-1)}{2} a_4 =  R_2
\end{equation}
Similarly, equating to zero the coefficient of $z^{n+1}$ in the expansion of $A_n$   yields
$$
 n b_2  + \frac{n(n-1)}{2} a_3 =  R_1.
$$
Finally, equating to zero the coefficient of $z^{n+1}$ in the expansion of $A_{n-1}$ yields
\begin{equation}\label{eq:second}
(n-1) b_3 + \frac{(n-1)(n-2)}{2} a_4 =  R_2
\end{equation}
Note that equations \eqref{eq:first} and \eqref{eq:second} together are equivalent to
$$
R_2 = \frac{n}{2} b_3 = - \frac{n(n-1)}{2} a_4.
$$

Finally, substituting these expressions into
equation \eqref{eq:identity}
  and comparing the coefficients of the monomials $z^i$ yields
	the system of ODEs for the functions for $g_i$.
\end{proof}

\subsection{Importance of the bounded state space}
In the last past of this section, we will be giving a detailed discussion of the importance of the assumption \ref{as:main}. Together with the assumption \ref{as:main} made in the beginning of this section, we will see that theorem \ref{th:main} has a financial impact. For technical issues, we start with the following lemma:

\begin{lemma} \label{th:lemma1}
Let $F: \RR^+ \times I \rightarrow \RR$ be a continuous function and $Z$ a stochastic process with state space $I = (\ell,r)$. If for any open interval $A \subseteq I$, there exists $0 <t_1 <t_2$ such that
$$
\PP(Z_t \in A \quad \forall t \in [t_1,t_2]) > 0
$$
if in addition for any fixed $T > 0$,
$$
F(T-t, Z_t(\omega)) = 0 \quad \text{for almost every $(t, \omega)$,}
$$
then
$$
F(x,z) = 0 \quad \forall (x,z) \in \RR^+ \times I
$$
\end{lemma}
\begin{proof}
suppose for contradiction that $F(x_0,z_0) \neq 0$ for some $(x_0, z_0) \in \RR^+ \times I$. Then by the continuity of $F$, there exits some rectangle $B$ surrounding $(x_0, z_0)$
$$
B = \{ (x,z)| x_1 \leq x \leq x_2; z_1 \leq z \leq z_2 \}
$$
such that $F(x,z) \neq 0$ for any $(x,z) \in B$. Now set the open interval $A = (z_1.z_2) \subseteq I$ and applying the condition of the lemma, there exists some $t_1 < t_2$ such that
$$
\PP \{ \omega : z_1 < Z_t(\omega) < z_2, \quad t_1 \leq t \leq t_2 \} > 0
$$
Now let $t_3 = \min \{ t_2, t_1+x_2-x_1 \}$ and fix $T = t_1 + x_2$. \\
Consider the function $F(t_1 + x_2 -\cdot, Z_\cdot (\cdot)) : [0, t_1 + x_2] \times \Omega \rightarrow \RR$ evaluating on the set $D$, where
$$
D = [t_1, t_3] \times \{ \omega : z_1 < Z_t(\omega) < z_2, \quad t_1 \leq t \leq t_2 \}
$$
Then the first argument of $F$ will be in range $[x_1, x_2]$ and the second argument will be in range $[z_1,z_2]$ by the definition of $D$. Hence we know that $F \neq 0$ on $D$ which has non-zero measure. Contradiction.
\end{proof}

With the above lemma, we can prove the following theorem that links the algebraic theorem \ref{th:main} to some financial aspects.
\begin{theorem} \label{th:linking}
Suppose there exists some bounded open interval $I \subseteq \RR$ and continuous functions $b:I \rightarrow \RR$, $\sigma:I \rightarrow \RR^*$ and $R:I \rightarrow \RR^+$ such that the solution $(Z_t)_{t \geq 0}$ to the SDE starting at any point $z \in I$
$$
dZ_t =  b(Z_t)dt + \sigma(Z_t)dW_t, \quad Z_0 = z \in I
$$
has the property that
$$
\PP(Z_t \in I) = 1, \forall t \geq 0
$$
and for any open interval $A \subseteq I$, there exists $0 <t_1 <t_2$ such that
$$
\PP(Z_t \in A, \quad \forall t \in [t_1,t_2]) > 0
$$
Then the function $H(x,z) : \RR^+ \times I \rightarrow \RR$ bounded on $K \times I$,where $K \subseteq \RR^+$ is any compact interval, is a solution to the PDE
$$
\partial_x H = b \partial_z H + \frac{1}{2}\sigma^2 \partial_{zz} H - RH
$$
subject to the boundary condition $H(0,z) = 1$ for all $z \in I$

\noindent
if and only if

the process $(M_t)_{t \leq T}$ defined by
$$
M_t = \exp{\left( -\int_0^t R(Z_s) ds\right)} H(T-t, Z_t)
$$
is a true martingale for each fixed $T > 0$, where $H \in C^{1,2}(\RR^+ \times I)$.
\end{theorem}

\begin{remark}
For any fixed $T >0$, the process $(M_t)_{t \leq T}$ is a true martingale is equivalent to the expression:
\begin{align*}
\EE \left[ e^{-\int_t^T R(Z_s)ds} | \mathcal F_t \right] = H(T-t, Z_t)
\end{align*}
The above formula has a financial interpretation. Indeed, if we identify the spot rate process $(r_t)_{t \geq 0}$ as $r_t = R(Z_t)$ and the bond price $P_t(T) = H(T-t, Z_t)$, then the above formula implies the bond price model admits no arbitrage in the sense of the first fundamental theorem of asset pricing.
\end{remark}

\begin{proof}
if the process $(M_t)_{t \leq T}$ is a true martingale for any fixed $T >0$, then by applying Ito's formula on $M_t$, we must have the drift term vanishing a.s. i.e.
$$
\partial_x H - b \partial_z H - \frac{1}{2}\sigma^2 \partial_{zz} H + RH = 0 \text{ evaluating at } (T-t,Z_t)
$$
By setting $F = \partial_x H - b \partial_z H - \frac{1}{2}\sigma^2 \partial_{zz} H + RH$ in the previous lemma, we conclude that $H$ is a solution to the PDE. i.e.
$$
\partial_x H - b \partial_z H - \frac{1}{2}\sigma^2 \partial_{zz} H + RH = 0 \text{ for all } (x,z) \in \RR^+ \times I
$$
Boundedness of $H$ follows by the fact that $R \geq 0$ and hence for any fixed $T >0$,
$$
0 \leq \EE \left[ e^{-\int_t^T R(Z_s)ds} | \mathcal F_t \right] = H(T-t, Z_t) \leq 1
$$

Conversely, if $H$ solves the PDE, then the process $(M_t)_{t \leq T}$ is a local martingale for any fixed $T >0$. Since $H$ is bounded and $R \geq 0$, then the process $(M_t)_{t \leq T}$ is a bounded local martingale and hence a true martingale.
\end{proof}
\begin{remark}
Actually the above theorem still holds whenever the spot rate function $R$ is bounded below instead of being strictly positive. And hence there exists polynomial models that are free of arbitrage but will allow possibly negative spot rate. However since allowing only non-negative spot rate is usually a desired property of an interest rate model, we will stick to the case where $R$ is non-negative.
\end{remark}

\begin{remark}
Consider the case $d=1$ and suppose that the function $H : \RR^+ \times I \to \RR$ takes the polynomial form of
equation \eqref{eq:poly}.  If $H$ is bounded, it must be the case that the state space $I \subseteq \RR$
of the factor process is bounded.  (However, notice that the boundedness of $H$ does not imply the boundedness
of $I$ in the case of exponential polynomial models.)
\end{remark}

The above theorem \ref{th:linking} suggests that we need to find factor process $Z$ that has bounded state space $I$. This can be achieved by applying Feller's test of explosion and hence there will be further constraints on the functions $b(z),a(z)$.

Before proceeding, we give a brief introduction to Feller's test. We fix an open interval $I = (\ell,r)$, where $\ell < r \in \RR$. Consider the SDE
$$
dZ_t = b(Z_t)dt + \sqrt{a(Z_t)}dW_t, \quad Z_0 = z \in I
$$
Where $b:I \rightarrow \RR$ and $a:I \rightarrow \RR^*$ are given measurable functions that satisfy the local integrability condition:
\begin{equation}\label{eq:locint}
\int_K \left( \frac{1}{a(z)} + \left| \frac{b(z)}{a(z)} \right| \right) dy < \infty, \quad \text{ for all compact } K \subset I
\end{equation}
Then there exists a unique weak solution until the explosion time $S$ defined as:
$$
S = \inf \{ t>0: X_t \notin (\ell,r) \}
$$
Feller's test function is defined as
$$
v(x) := \int_c^x \int_z^x \frac{1}{a(z)} e^{\int_y^z \frac{2b(w)}{a(w)}dw} dy dz, \quad x\in I
$$
For some fixed constant $c \in I$.
\begin{theorem}[Feller's test of explosion]
With the above assumptions,
$$
\PP(S = \infty) = 1 \Leftrightarrow \lim_{x \downarrow \ell}v(x) = \infty = \lim_{x \uparrow r} v(x)
$$
and the finiteness of the Feller's test function $v(x)$ is independent of the choice of constant $c$.
\end{theorem}
\begin{lemma}
In polynomial models, if the factor process $(Z_t)_{t \geq 0}$ is non-explosive and has a bounded state space $I$. Then $a(z)$ must has at least two different real roots.
\end{lemma}
\begin{proof}
In polynomial models, the functions $b(z),a(z)$ must be polynomials and hence local integrability condition \eqref{eq:locint} holds on some bounded interval $I$.
Suppose for contradiction that $a(z)$ has no real root. Then $a(z) \neq 0$ for any $z\in \RR$. Therefore the Feller's test function $v(x)$ will be finite for all $x\in \RR$.

If $a(z)$ has only one real root $d$, then $v(x)$ will be finite for all $x \neq d$. Therefore in both ways we cannot find two distinct real numbers $\ell,r$ such that $\lim_{x \downarrow \ell}v(x) = \infty = \lim_{x \uparrow r} v(x)$. By Feller's test, the state space $I$, if exists, must be unbounded.
\end{proof}

\subsection{Polynomial SDEs with a unique bounded solution}
With the discussions so far, we may appreciate the role played by bounded factor process $Z$ in polynomial models. To be more specific, the algebraic theorem \ref{th:main} allows us to formally write down a candidate SDE for the factor process $Z$ that will possibly lead to a polynomial type of model. Thus if the SDE happens to have a solution (not necessarily bounded), then the following model is free of arbitrage:
\begin{align*}
\text{spot rate process } r_t &:= R(Z_t) \\
\text{time $t$ price of bond with maturity $T$ } P_t(T)&:= \sum_{k=0}^n g_k(T-t) Z_t^k
\end{align*}
simply because the discounted bond price is a local martingale for every maturity date $T > 0$. In particular, if the SDE has a bounded solution then the above model is not only free of arbitrage but also allows the following pricing identity:
$$
P_t(T):= \sum_{k=0}^n g_k(T-t) Z_t^k = \EE \left[ e^{-\int_t^T R(Z_s)ds} | \mathcal F_t \right]
$$

Therefore we are urged to determine if the candidate SDE we get from applying the algebraic theorem \ref{th:main} allows such pricing formula. And the following theorem does the job. It characterize SDEs with polynomial drift and square of volatility that has a unique solution in a bounded state space.
\begin{theorem}[Characterisation of SDE with polynomial coefficients that has a unique bounded solution] \label{th:polyf}
Suppose $a(z)$ and $b(z)$ are polynomials, with $a(\ell)=a(r) = 0$ and
$a(z) > 0$ for $\ell < z < r $.

Let
$$
D(z) := 2b(z) + a'(z) h(a'(z),b(z))
$$
where
\begin{align*}
h(x,y) =  \left\{ \begin{array}{cc}
                 1, & x=y=0 \\
                 -1, &  \mbox{otherwise}
               \end{array} \right.
\end{align*}
Then the following are equivalent:

(1) For every $z \in (\ell, r)$ the SDE
$$
dZ_t = b(Z_t) + \sqrt{a(Z_t)} dW_t
$$
has a unique strong solution valued in $(\ell, r)$ such that  $Z_0= z$.

(2)  $D(\ell) \ge 0 \ge D(r)$.
\end{theorem}

Before proving the above theorem, we will find the following lemma useful.

\begin{lemma}
Suppose $a(z)$ and $b(z)$ are polynomials, with $a(r) = 0$ and
$a(z) > 0$ on the interval $c < z < r $ for some constant $c$. Suppose we write $a(z) = A(z)(r-z)^\alpha$ and $b(z) = B(z)(r-z)^\beta$ for some integers $\alpha \geq 1$, $\beta \geq 0$ such that $A(z)$, $B(z)$ are polynomials with $A(r) \neq 0 \neq B(r)$. Define $T = \frac{2B(r)}{A(r)} \in \RR \backslash \{0 \}$ and
$$
v(x) = \int_c^x \int_z^x \frac{1}{a(z)} e^{\int_y^z \frac{2b(w)}{a(w)}dw} dy dz
$$
Then the following statements hold:

\noindent
(i) If $\alpha = 1, \beta = 0$, then $\lim_{x \uparrow r} v(x) = \infty$ if and only if $T \leq -1$

\noindent
(ii) If $\alpha = 1, \beta \geq 1$, then $\lim_{x \uparrow r} v(x) < \infty$

\noindent
(iii) If $\alpha \geq 2, \beta = 0$, then $\lim_{x \uparrow r} v(x) = \infty$ if and only if $T < 0$

\noindent
(iv) If $\alpha \geq 2, \beta \geq 1$, then $\lim_{x \uparrow r} v(x) = \infty$
\end{lemma}
\begin{proof}
Before proceeding to the actual proof, we first define the notation
\begin{align*}
f(x) \sim g(x) \text{ as } x \rightarrow c
\end{align*}
as
\begin{align*}
\lim_{x \rightarrow c} \frac{f(x)}{g(x)} = 1
\end{align*}

Also notice that the integral
$$
v(r) = \int_c^r \int_z^r \frac{1}{a(z)} e^{\int_y^z \frac{2b(w)}{a(w)}dw} dy dz
$$
is well defined with possible value $\infty$ since the integrand is non-negative.

First we consider the case $\alpha = 1$, there are two subcases to consider.

\noindent
Subcase 1: $\beta \geq 1$

In this case, we have:
$$
\frac{2b(w)}{a(w)} =\frac{2B(w)}{A(w)} (r-w)^{\beta - 1}
$$
and there is no singularity around the point $w=r$, hence the function
$$
e^{\int_y^z \frac{2b(w)}{a(w)}dw}
$$
is continuous and bounded on the interval $[c,r]$. Therefore we have:
$$
v(r) = \int_c^r \int_z^r \frac{1}{a(z)} e^{\int_y^z \frac{2b(w)}{a(w)}dw} dy dz < \infty
$$
simply because $\int_c^r \int_z^r \frac{1}{a(z)} dydz < \infty$.

\vspace{8pt}
\noindent
Subcase 2: $\beta = 0$

We first set
$$
e(w) := \frac{2b(w)}{a(w)} - T (r-w)^{-1}
$$
Then $e(w)$ is a continuous function on the interval $[c,r)$. Especially, when $w \uparrow r$, we can apply L'Hospitals rule here to give:
\begin{align*}
\lim_{w \uparrow r} e(w) &= \lim_{w \uparrow r} \frac{\frac{2B(w)}{A(w)}-T}{r-w} \\
                         &= 2 \cdot \frac{B'(r)A(r) - B(r)A'(r)}{A^2(r)} < \infty \text{ since $A(r) \neq 0$}
\end{align*}
Then $e(w)$ is also bounded on the interval $[c,r)$ and we then have the following estimates:
\begin{align*}
\int_y^z \frac{2b(w)}{a(w)} dw &=\int_y^z T(r-w)^{-1} + e(w) dw \\
     &=T \cdot \log{\left( \frac{r-y}{r-z} \right)} + E(y,z)
\end{align*}
where the error term $E(y,z)$ is bounded as $y,z \in [c,r)$. Therefore by mean value theorem for integrals, we get:
\begin{align*}
v(r) &= \int_c^r \int_z^r \frac{1}{a(z)} e^{\int_y^z \frac{2b(w)}{a(w)}dw} dy dz \\
     &= \int_c^r \frac{1}{A(z)(r-z)} \int_z^r \left( \frac{r-y}{r-z} \right)^T \cdot e^{E(y,z)} dy dz \\
     &= \int_c^r \frac{1}{A(z)} (r - z)^{-1-T} \cdot e^{E(\eta(z),z)} \int_z^r (r - y)^T dy dz
\end{align*}
where $\eta(z)$ is some constant in $(z,r)$. It is clear that the finiteness of $v(r)$ depends on the integral
$$
\int_c^r (r - z)^{-1-T} \int_z^r (r - y)^T dy dz
$$
simply because the remaining part of the integrand $A^{-1}(z) e^{E(\eta(z),z)}$ is finite over the integrating interval $[c,r)$. It's clear that the above integral is infinite if and only if $T \leq -1$.

Hence $v(r) = \infty$ if and only if $T \leq -1$.

\vspace{15pt}
\noindent
For the case when $\alpha \geq 2$, there are three subcases to consider.

\noindent
Subcase 1: $\beta \geq \alpha$

In this case, we have:
$$
\frac{2b(w)}{a(w)} =\frac{2B(w)}{A(w)} (r-w)^{\beta - \alpha}
$$
and there is no singularity around the point $w=r$, hence the function
$$
e^{\int_y^z \frac{2b(w)}{a(w)}dw}
$$
is continuous and bounded on the interval $[c,r]$. Therefore we have:
$$
v(r) = \int_c^r \int_z^r \frac{1}{a(z)} e^{\int_y^z \frac{2b(w)}{a(w)}dw} dy dz = \infty
$$
simply because $\int_c^r \int_z^r \frac{1}{a(z)} dydz = \infty$.

\vspace{8pt}
\noindent
Subcase 2: $\beta = \alpha -1$

In this case, we have as $w \uparrow r$,
$$
\frac{2b(w)}{a(w)} \sim  T(r-w)^{\beta - \alpha}
$$
then for any $\epsilon > 0$, when $y,z$ are closed to $r$, we have
$$
(1-\epsilon) T \int_y^z (r - w)^{\beta - \alpha} dw < \int_y^z \frac{2b(w)}{a(w)} dw < (1+\epsilon) T \int_y^z (r - w)^{\beta - \alpha} dw
$$
or
$$
(1+\epsilon) T \int_y^z (r - w)^{\beta - \alpha} dw < \int_y^z \frac{2b(w)}{a(w)} dw < (1-\epsilon) T \int_y^z (r - w)^{\beta - \alpha} dw
$$
depending on the sign of $T$.
Therefore when picking the constant $c$ arbitrarily close to $r$, we must have the value of $v(r)$ is between $u((1-\epsilon)T)$ and $u((1+\epsilon)T)$, where $u(t)$ is defined as
$$
u(t) := \int_c^r \int_z^r \frac{1}{a(z)} e^{t \int_y^z (r - w)^{\beta - \alpha}dw} dy dz
$$
For the subcase $\beta - \alpha = -1$, we have:
\begin{align*}
u(t) &= \int_c^r \int_z^r \frac{1}{A(z)(r-z)^\alpha} (r-y)^t (r-z)^{-t} dy dz \\
     &= \int_c^r  \frac{1}{A(z)(r-z)^{\alpha + t}} \int_z^r (r-y)^t dy dz
\end{align*}
when $t \leq -1$, $u(t) = \infty$ since
$$
\int_z^r (r-y)^t = \infty
$$
On the other hand, when $t > -1$,
\begin{align*}
u(t) &= \int_c^r  \frac{1}{A(z)}(r-z)^{-\alpha -t} \cdot \frac{1}{t+1} (r-z)^{t+1} dz \\
     &= \int_c^r \frac{1}{(t+1)A(z)}(r-z)^{1-\alpha} dz = \infty \text{ since $\alpha \geq 2$}
\end{align*}
Hence $v(r) = \infty$ since $u((1-\epsilon)T)$ and $u((1+\epsilon)T)$ are both $\infty$.

\vspace{8pt}
\noindent
Subcase 3: $0 \leq \beta \leq \alpha - 2$

In this case, we again have the value of $v(r)$ is between $u((1-\epsilon)T)$ and $u((1+\epsilon)T)$, where $u(t)$ is defined in subcase 2. Now since $\beta - \alpha \leq -2$, $u(t)$ takes the form as:
\begin{align*}
u(t) &= \int_c^r \int_z^r \frac{1}{A(z)(r-z)^\alpha} e^{\frac{t}{\beta - \alpha + 1}\left( (r-y)^{\beta - \alpha + 1} - (r-z)^{\beta - \alpha +1} \right) } dy dz \\
     &= \int_c^r \frac{1}{A(z)(r-z)^\alpha} e^{-\frac{t}{\beta - \alpha + 1} (r-z)^{\beta - \alpha +1}} \int_z^r  e^{\frac{t}{\beta - \alpha + 1} (r-y)^{\beta - \alpha + 1}} dy dz
\end{align*}

if $t < 0$ then $\frac{t}{\beta - \alpha + 1} > 0$ and hence
\begin{align*}
\int_z^r  e^{\frac{t}{\beta - \alpha + 1} (r-y)^{\beta - \alpha + 1}} dy &> e^{\frac{t}{\beta - \alpha + 1} (r-z)^{\beta - \alpha + 1}} \int_z^r dy \\
                                                                         &= e^{\frac{t}{\beta - \alpha + 1} (r-z)^{\beta - \alpha + 1}} (r - z)
\end{align*}
and
$$
u(t) > \int_c^r \frac{1}{A(z)(r-z)^\alpha} (r - z) dz = \infty
$$
and hence $v(r) = \infty$ when $T < 0$ since both $u((1-\epsilon)T)$ and $u((1+\epsilon)T)$ are $\infty$ in this case.

if $t > 0$, let $\frac{f(z)}{A(z)}$ be the integrand of $u(t)$. i.e.
$$
f(z) = \frac{1}{(r-z)^\alpha} e^{-\frac{t}{\beta - \alpha + 1} (r-z)^{\beta - \alpha +1}} \int_z^r  e^{\frac{t}{\beta - \alpha + 1} (r-y)^{\beta - \alpha + 1}} dy
$$
then for any $-1 \leq \gamma < 0$, we have
$$
\frac{f(z)}{(r-z)^\gamma} = \frac{\int_z^r  e^{\frac{t}{\beta - \alpha + 1} (r-y)^{\beta - \alpha + 1}} dy}{(r-z)^{\alpha + \gamma}e^{\frac{t}{\beta - \alpha + 1} (r-z)^{\beta - \alpha +1}}}
$$
Recall that we have the conditions $\alpha \geq 2$, $\beta - \alpha \leq -2$, $-1 \leq \gamma < 0$ and $t > 0$. Therefore as $z \uparrow r$ both the numerator and denominator tends to 0 and we can apply L'Hospitals rule and get
$$
\lim_{z \uparrow r} \frac{f(z)}{(r-z)^\gamma} = \lim_{z \uparrow r} \frac{1}{(\alpha + \gamma)(r-z)^{\alpha + \gamma - 1} + t(r-z)^{\beta + \gamma}}
$$
Hence when $\beta = 0$, we can take $\gamma = -\tfrac{1}{2}$ to conclude that $\lim_{z \uparrow r} \frac{f(z)}{(r-z)^\gamma} = 0$. Then
$$
u(t) = \int_c^r f(z) dz < \int_c^r (r-z)^{-\tfrac{1}{2}} dz < \infty \text{ when $c$ is closed to $r$.}
$$
and hence $v(r) < \infty$ when $T > 0$ and $\beta = 0$, since both $u((1-\epsilon)T)$ and $u((1+\epsilon)T)$ are finite in this case.

Similarly when $\beta \geq 1$, we can take $\gamma = -1$ to conclude that $\lim_{z \uparrow r} \frac{f(z)}{(r-z)^\gamma} = \infty$. Then
$$
u(t) = \int_c^r f(z) dz > \int_c^r (r-z)^{-1} dz = \infty \text{ when $c$ is closed to $r$.}
$$
and hence $v(r) = \infty$ when $T > 0$ and $\beta \geq 1$, since both $u((1-\epsilon)T)$ and $u((1+\epsilon)T)$ are $\infty$ in this case.

\end{proof}

\begin{remark}
For the other limit point $\ell$, we may redefine $a(z) = A(z)(z-\ell)^\alpha$, $b(z) = B(z)(z-\ell)^\beta$ and $T = \frac{2B(\ell)}{A(\ell)}$. The conclusions are the same when we replace $T \leq -1$ and $T < 0$ by $T \geq 1$ and $T > 0$.
\end{remark}

Now we are ready to prove theorem \ref{th:polyf}.
\begin{proof}
By applying the theorem of Feller's test of explosion, we are left to show that
$$
D(\ell) \geq 0 \geq D(r)
$$
if and only if
$$
\lim_{x \downarrow \ell} v(x) = \infty = \lim_{x \uparrow r} v(x)
$$
where $v(x)$ is the Feller's test function. We will consider the upper limit $r$, the other limit $\ell$ is similar. By using the previous lemma, it suffices to show that $D(r) \leq 0$ if and only if one of the cases from (i) to (iv) in the previous lemma holds, which can be checked case by case.

\end{proof}

\subsection{Conclusion: an algorithm}
In conclusion the algorithm of searching for polynomial models may be described as follows:

\noindent
Step 1: We impose constraints on $b,\sigma$ by using theorem \ref{th:main} to find a candidate SDE
$$
dZ_t = b(Z_t)dt + \sigma(Z_t)dW_t
$$
that the factor process must satisfy. By this stage, there is no guarantee that the SDE has a solution.

\vspace{8pt}
\noindent
Step 2: Then we apply theorem \ref{th:polyf} to impose further constraints on the functions $b,\sigma$ so that the SDE has a unique bounded non-explosive solution. Theorem \ref{th:linking} ensures the model we are considering will be free of arbitrage in the sense that all discounted zero-coupon bond prices are true martingales.

\vspace{8pt}
\noindent
Step 3: Revoke theorem \ref{th:main} again to solve the coefficient functions $(g_i)_{i}$ and hence the polynomial model completely.
\begin{remark} \label{re:tech}
Careful readers may notice that in step 2, we didn't check the condition in theorem \ref{th:linking}, namely for any open interval $A \subseteq I$, there exists $0 <t_1 <t_2$ such that
$$
\PP(Z_t \in A, \quad \forall t \in [t_1,t_2]) > 0
$$

However this condition holds by a simple application of Feller's test. Indeed in the case of polynomial model, the functions $b(z),a(z)$ are polynomials. Take any $A = (m,n) \subseteq (\ell,r) = I$, the local integrability condition on functions $\frac{1}{a(z)}$ and $\frac{b(z)}{a(z)}$ implies that the Feller's test function $v(x)$ evaluating at points $x = m,n$ are finite. Hence by applying Feller's test
$$
\PP(T_m < \infty) = 1 = \PP(T_n < \infty)
$$
where $T_m, T_n$ are the hitting times of point $m,n$. Without loss of generality assume that initially, the process $Z$ starts at point $Z_0 = z < m$. Then by the continuity of the process $Z$, we must conclude that
$$
\PP( T_m < T_n < \infty) = 1
$$
Hence there exist some $t_1 < t_2$ such that
$$
\PP( T_m < t_1 < t_2 < T_n < \infty) > 0
$$
which is exactly the condition we want.
\end{remark}
\section{Two examples} \label{se:examples}
In this section, we will consider two explicit families of parametric models. Both of them are quadratic models corresponding to $n = 2$. We will solve them and try to calibrate the parameters by using the US Treasury rates from 2006 to 2014. We get the data from yahoo finance and table \ref{tab:yahoo} shows some of the raw data. The data are sampled weekly with eleven different time to maturity.

\begin{table}[h]
\centering
\begin{tabular}{l|l|l|l|l|l|l|l|l|l|l|l}
Date  & 1M  & 3M  & 6M & 1Y & 2Y & 3Y & 5Y & 7Y & 10Y & 20Y & 30Y \\
\hline
060210&4.33 &4.5  & 4.68 & 4.67 & 4.64 & 4.61 & 4.54 & 4.55 & 4.56 & 4.73 & 4.53 \\
060217&4.39 &4.55 & 4.7  & 4.7  & 4.69 & 4.67 & 4.59 & 4.58 & 4.59 & 4.76 & 4.56 \\
\vdots&\vdots&\vdots&\vdots&\vdots&\vdots&\vdots&\vdots&\vdots&\vdots&\vdots&\vdots\\
140502&0.01 &0.03  & 0.05 & 0.1  & 0.43 & 0.89 & 1.7  & 2.25 & 2.66 & 3.2  & 3.44 \\
140509&0.02 &0.03  & 0.05 & 0.1  & 0.41 & 0.89 & 1.65 & 2.19 & 2.62 & 3.15 & 3.42 \\
\end{tabular}
\caption{US Treasury rates from 2006 June 10th to 2014 May 9th, sampling weekly with time to maturity
          ranging from 1 month to 30 years. There are a total of $430 \times 11$ observations. The numbers in the table are in percentage. The full table contains a total number of 430 sample dates. Here we just provide the data of the first and last two sample dates.}
\label{tab:yahoo}
\end{table}

In general, given any parametric model, let $y_i(x)$ be the time $i$ yield with time to maturity $x$ calculated from the model. We can calibrate the model parameters by minimising the sum of squares of the observed yield from the data. To be more specific, let $Y_i(x_j)$ denote the observed yield at i-th sample date with time to maturity $x_j$. We want to choose models parameters such that the error defined below is small:
\begin{align*}
E   &:= \sum_{i,j} (Y_i(x_j) - y_i(x_j))^2\\
\end{align*}

In practice, suppose the model parameters $\lambda$ take values in the parameter space $\Lambda \subseteq \RR^k$, we can start by choosing any $\lambda_0, \lambda_1 \in \Lambda$ and calculate the corresponding $E(\lambda_0), E(\lambda_1)$. We can then compare $E(\lambda_0)$ and $E(\lambda_1)$ and keep $\lambda_1$ only if $E(\lambda_1) < E(\lambda_0)$. Otherwise we drop $\lambda_1$ and try a new candidate $\lambda_2$ chosen randomly in the parameter space $\Lambda$. A program can be written to perform the above algorithm many times and the resulting values for the parameters can fit the data very well.

We will now introduce the first example and then use the algorithm described above to fit the data.
\begin{ex}[Four-parameter family]
Here the spot rate process $r$ solves the SDE
\begin{align*}
dr_t &= \alpha (\beta -r_t) dt + \sqrt{r_t(k-r_t)(\ell-r_t)}dW_t
\end{align*}
with parameter $\alpha >0$ and $0< \beta < k < \ell$.   Roughly speaking, the
dynamics of interest rate in this model resemble the Cox--Ingersoll--Ross
process when $r_t$ is very small.   The parameters $\beta$ intuitively
plays the role of a long time mean level, while $\alpha$
controls the speed of mean reversion.
 However, in this model, the interest rate
stays within the bounded interval $I = (0,k)$.

Indeed, theorem \ref{th:polyf} shows that
$$
\PP( 0 < r_t < k \mbox{ for all } t \ge 0) = 1
$$
as long as the initial condition $r_0$ is in $(0,k)$ and
$$
D(0) \geq 0 \geq D(k)
$$
where
$$
D(r) = 2\alpha (\beta - r) - \left( (k-z)(l-z) - z(l-z) - z(k-z) \right)
$$
i.e.
$$
\frac{\alpha \beta}{kl} \geq \frac{1}{2}  \mbox{ and } \frac{ \alpha (k-\beta)}{k (\ell-k)} \geq \frac{1}{2}
$$
Notice that this
is a quadratic $n=2$ model.   The corresponding matrix $S$ of this family takes the form
\begin{align*}
S = \left(
  \begin{array}{ccc}
    0 & \alpha \beta & 0 \\
    -1 & -\alpha & 2 \alpha \beta +k\ell \\
    0 & -1 & -2\alpha -k-\ell \\
  \end{array}
\right)
\end{align*}
from which the function $G = (g_0, g_1, g_2)^\top$ can be calculated by solving the ODE $\dot G = S G$
subject to $G(0) = (1, 0, 0)^\top$.

Notice that the process is ergodic in the pricing measure $\QQ$, and its invariant density is given by
the unique stationary solution of the corresponding Fokker--Planck PDE:
\begin{align*}
f(r) & = \frac{C}{\sigma(r)^2} e^{ \int_{r_0}^r \frac{2 b(\rho)}{\sigma(\rho)^2} d\rho} \\
& \propto r^{2\zeta - 1} (k-r)^{2 \eta -1}  (\ell-r)^{-2\theta-1}
\end{align*}
where
$$
\zeta = \frac{\alpha \beta}{k\ell}, \ \ \eta = \frac{ \alpha (k-\beta)}{k (\ell-k)},
\ \ \theta = \frac{ \alpha (\ell-\beta)}{\ell (\ell-k)}
$$
and where $C > 0$ is such that $\int_0^k f(r) dr = 1$.

The characteristic polynomial of matrix $S$ is given by:
$$
\chi(\lambda) = \lambda^3 + (3\alpha +k+l)\lambda^2 + (\alpha(2\alpha +k+l)+3\alpha \beta +kl)\lambda + \alpha \beta(2\alpha +k+l)
$$
Notice that
\begin{align*}
\chi(0) &= \alpha \beta(2\alpha +k+l) & >0 \\
\chi(-\beta) &= \beta(k-\beta)(\beta -l) & <0\\
\chi(-(\alpha +k)) &= (\alpha +k)(\alpha k -3\alpha \beta) + \alpha \beta(2\alpha +k +l)\\
                &\geq (\alpha +k)(\alpha k -3\alpha \beta) + 2\alpha \beta(\alpha +k)\\
                &= \alpha(\alpha+k)(k-\beta) & >0\\
\chi(-(2\alpha+k+l)) &= -2\alpha \beta -kl &<0
\end{align*}
Hence the equation $\chi(\lambda) = 0$ has three distinct negative roots. Therefore matrix $S$ will always has three negative eigenvalues $\lambda_1,\lambda_2,\lambda_3$ such that:
$$
-(2\alpha +k +l) < \lambda_3 < -(\alpha +k) < \lambda_2 < -\beta < \lambda_1 < 0
$$
and by remark \ref{re:pfun} the resulting bond prices may always be interpreted as a linear combination of bond prices from models with constant positive interest rates $r = -\lambda_i$.

On the other hand, given initial spot rate $r_0$, the time 0 bond price and yield with time to maturity $x$ can be expressed as:
\begin{align*}
P(x,r_0) &= g_0(x) + g_1(x)r_0 + g_2(x) r_0^2\\
y(x,r_0) &= -\frac{\log{P(x,r_0)}}{x}
\end{align*}
Hence in this model, we can calculate the whole yield curve once we know the current spot rate $r_0$. In order to calibrate the parameters $\alpha, \beta, k ,l$ by using data from table \ref{tab:yahoo}, we can use the one month yield as an approximation to the spot rate $r_0$ and try to fit the rest of the data by using least square principle.

By performing 2,000 random searches over the region $\alpha \in (0,1)$, $\beta \in (0, 0.1)$, $k \in (0,0.2)$ and $l \in (0,0.3)$, the parameters $\beta = 0.03, \alpha = 0.5, k=0.1, l=0.2$ fit reasonably well. The corresponding error is $E = 0.3246$. Since there are 4,300 terms in the expression of $E$, we can say that the average difference is then
$$
\text{average difference\%} = \sqrt{0.3246/4300} \times 100\% = 0.87\%
$$
Notice that this choice of parameters satisfies the condition of theorem \ref{th:polyf}, therefore the corresponding spot rate process will not hit the boundary. The corresponding matrix is given by
\begin{align*}
S = \left(
      \begin{array}{ccc}
        0 & 0.015 & 0 \\
        -1 & -0.5 & 0.05 \\
        0 & -1 & -1.3 \\
      \end{array}
    \right).
\end{align*}
With the above parameter values, we can simulate the spot rate process $r$.
A typical sample path with initial a very low spot rate $r_0 = 10^{-4}$ in line with current market conditions, is illustrated
in Figure \ref{fi:spot}.

The initial yield curve, calculated by the formula
$$
y_0(T) = - \frac{1}{T} \log H(T, r_0),
$$
is given in Figure \ref{fi:yield}
\begin{figure}
\centering
\includegraphics[scale = 0.3]{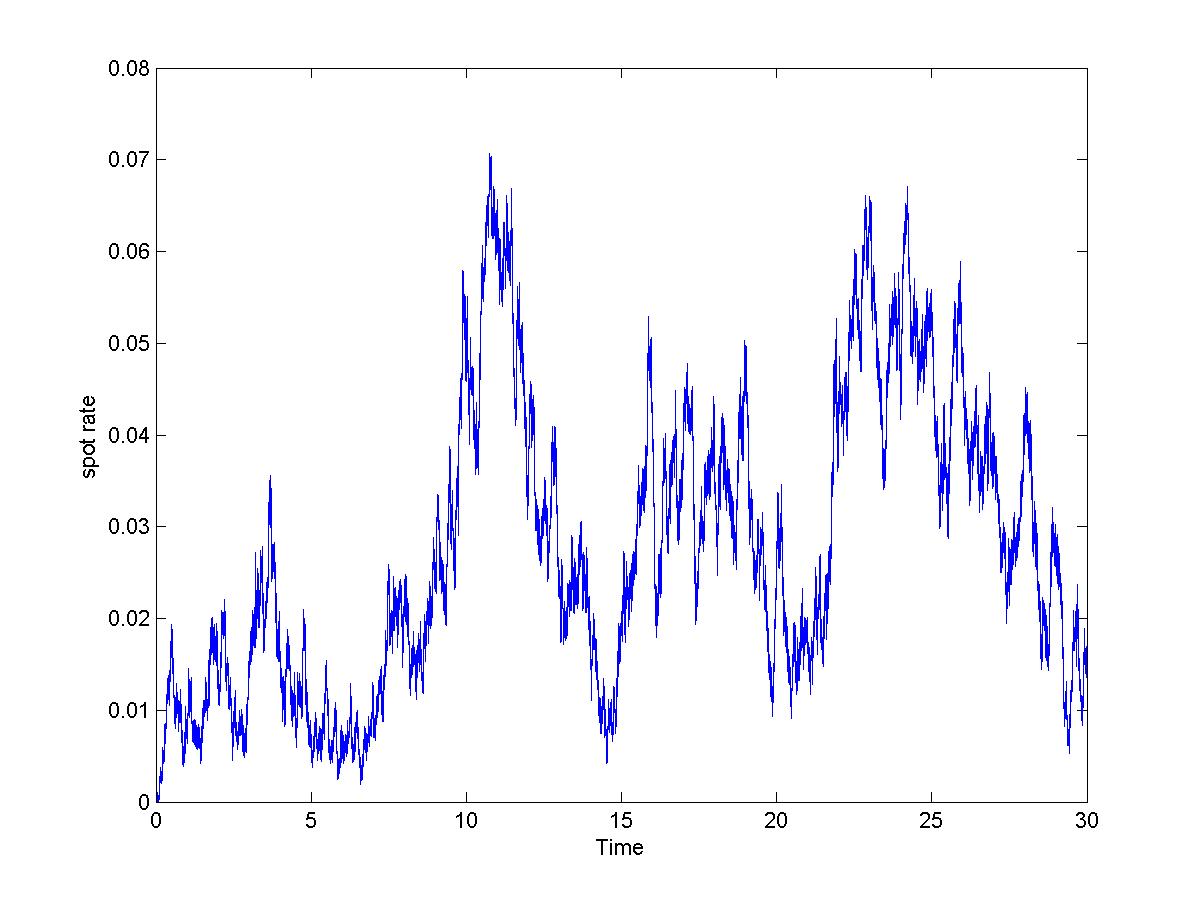}
\caption{Typical sample path of the spot rate process $r$, where $dr_t = \alpha (\beta -r_t) dt + \sqrt{r_t(k-r_t)(\ell-r_t)}dW_t$ with model parameters $\alpha = 0.5, \beta = 0.03, k = 0.1, l = 0.2$. The initial spot rate is set to $r_0 = 0.01\%$.}
\label{fi:spot}
\end{figure}
\begin{figure}
\centering
\includegraphics[scale = 0.25]{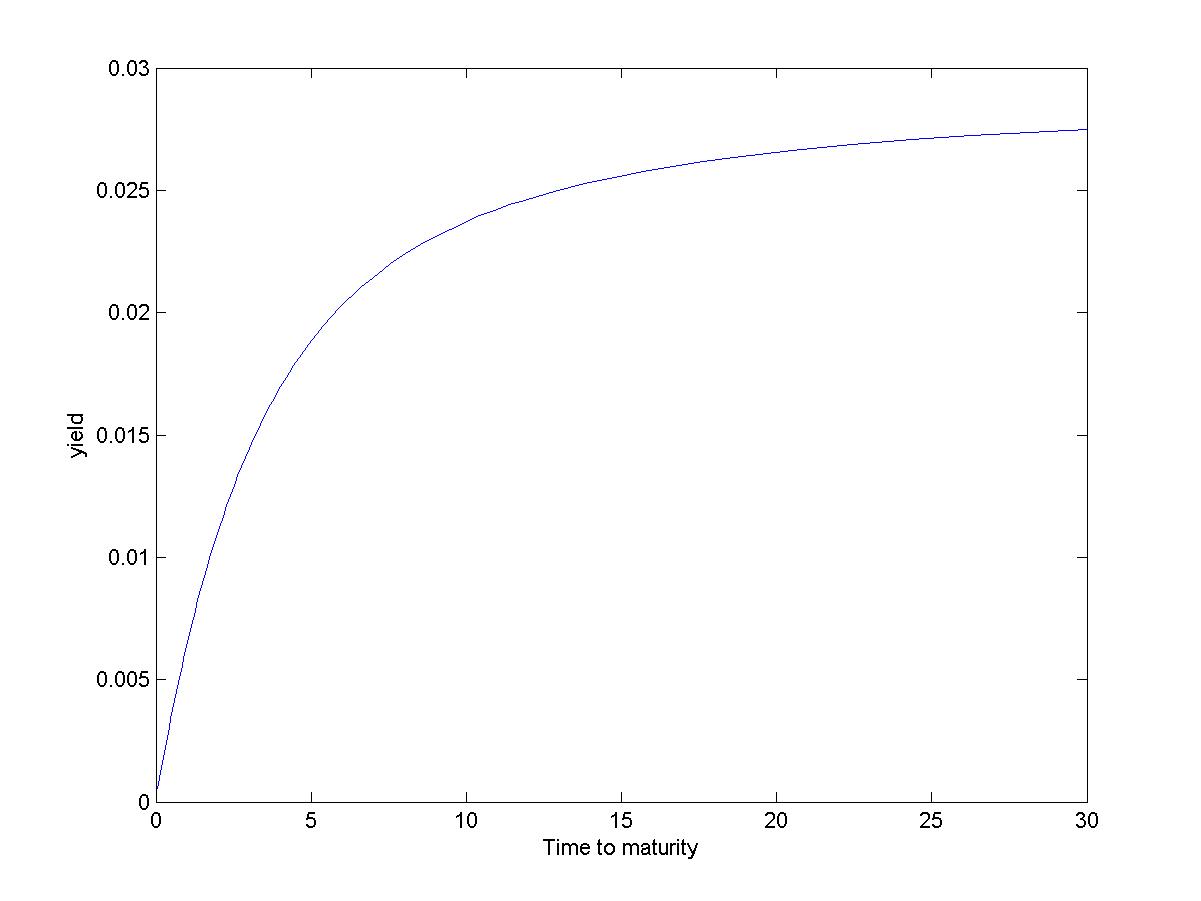}
\caption{Yield curve with model parameters $\alpha = 0.5, \beta = 0.03, k = 0.1, l = 0.2$ and initial spot rate $r_0 = 0.01\%$.}
\label{fi:yield}
\end{figure}

By changing the initial spot rate $r_0$, we can also get different shapes of yield curve as shown
in Figure \ref{fi:yield2}.
\begin{figure}
\centering
\includegraphics[scale = 0.25]{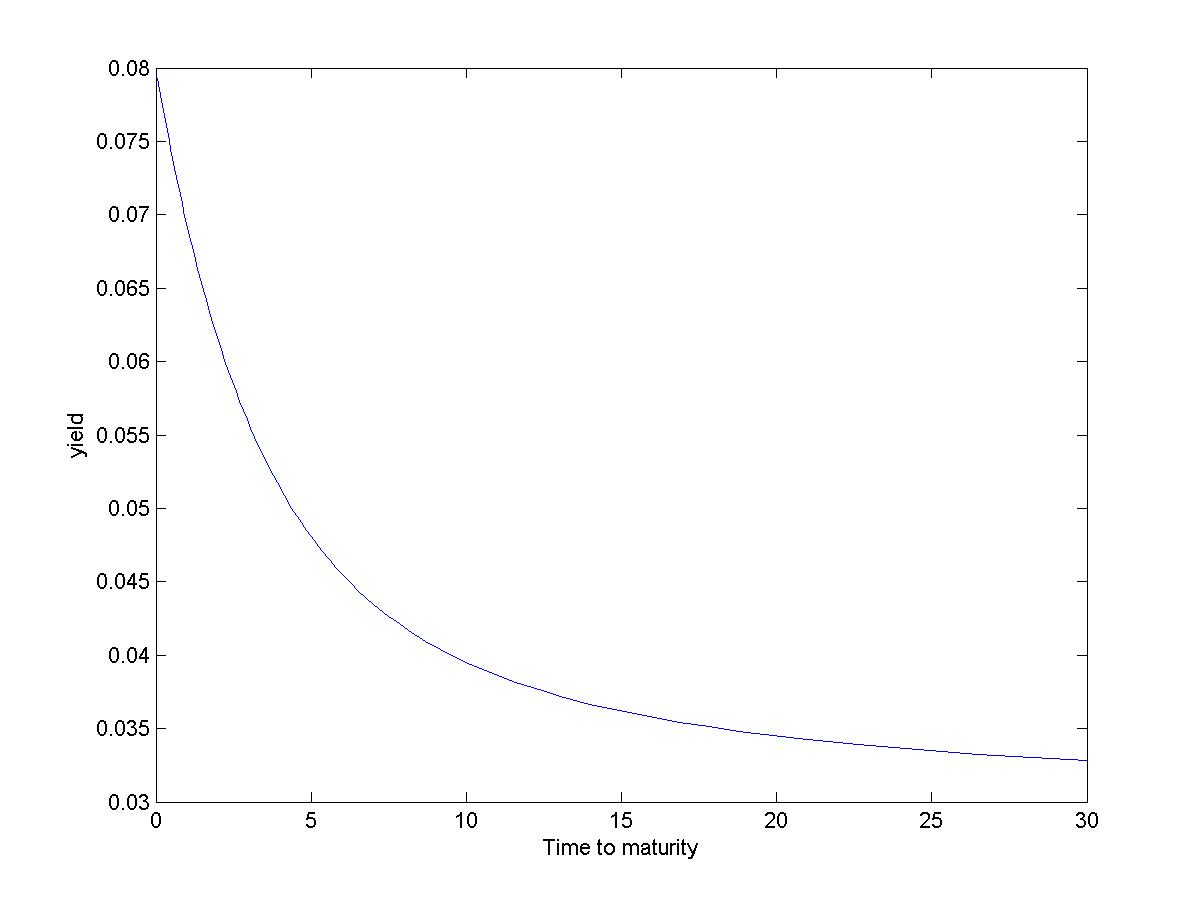}
\caption{Yield curve with model parameters $\alpha = 0.5, \beta = 0.03, k = 0.1, l = 0.2$ and initial spot rate $r_0 = 8\%$.}
\label{fi:yield2}
\end{figure}

The graphs of the functions $g_i$ and $P_i$ function are shown in Figures \ref{fi:g} and \ref{fi:P}
where
$$
H(x,r) = \sum_{k=0}^2 g_k(x) r^k = \sum_{i=0}^2 P_i(r) e^{\lambda_i x}
$$
where $\lambda_i \in \{ -0.0294, -0.5377, -1.2329\}$ are the eigenvalues of $S$.
\begin{figure}
\centering
\begin{subfigure}[b]{0.49\textwidth}
\centering
\includegraphics[width = \textwidth]{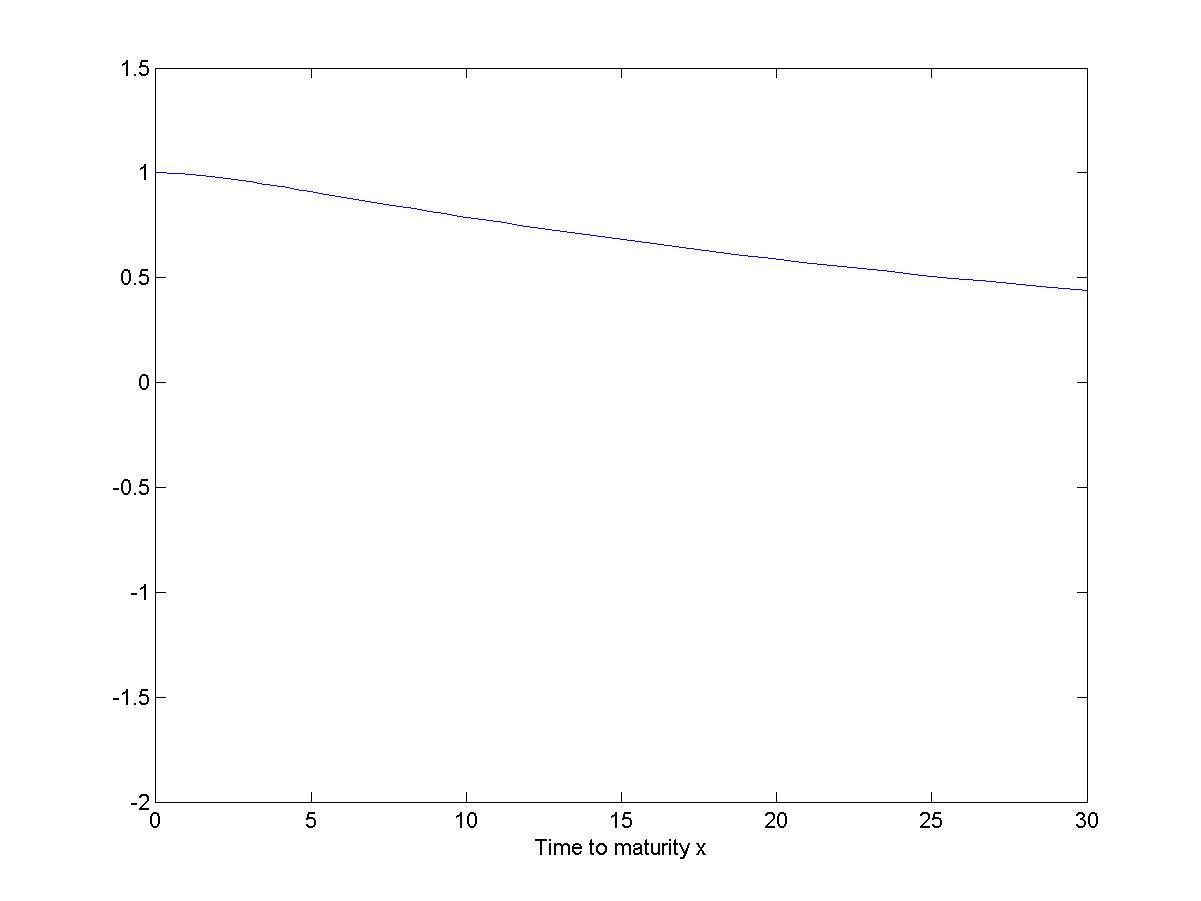}
\caption{graph of $g_0(x)$}
\end{subfigure}
\hfill
\begin{subfigure}[b]{0.49\textwidth}
\centering
\includegraphics[width = \textwidth]{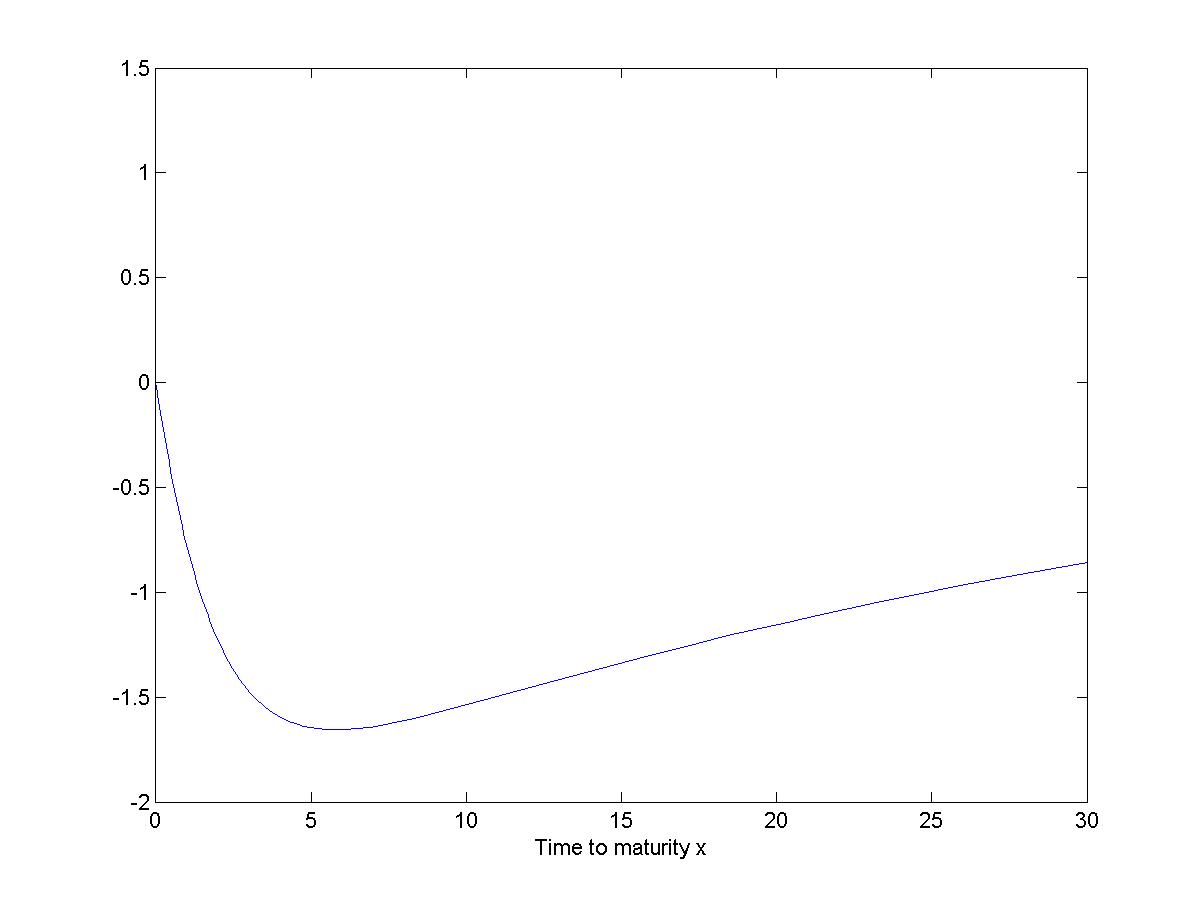}
\caption{graph of $g_1(x)$}
\end{subfigure}
\hfill
\begin{subfigure}[b]{0.5\textwidth}
\centering
\includegraphics[width = \textwidth]{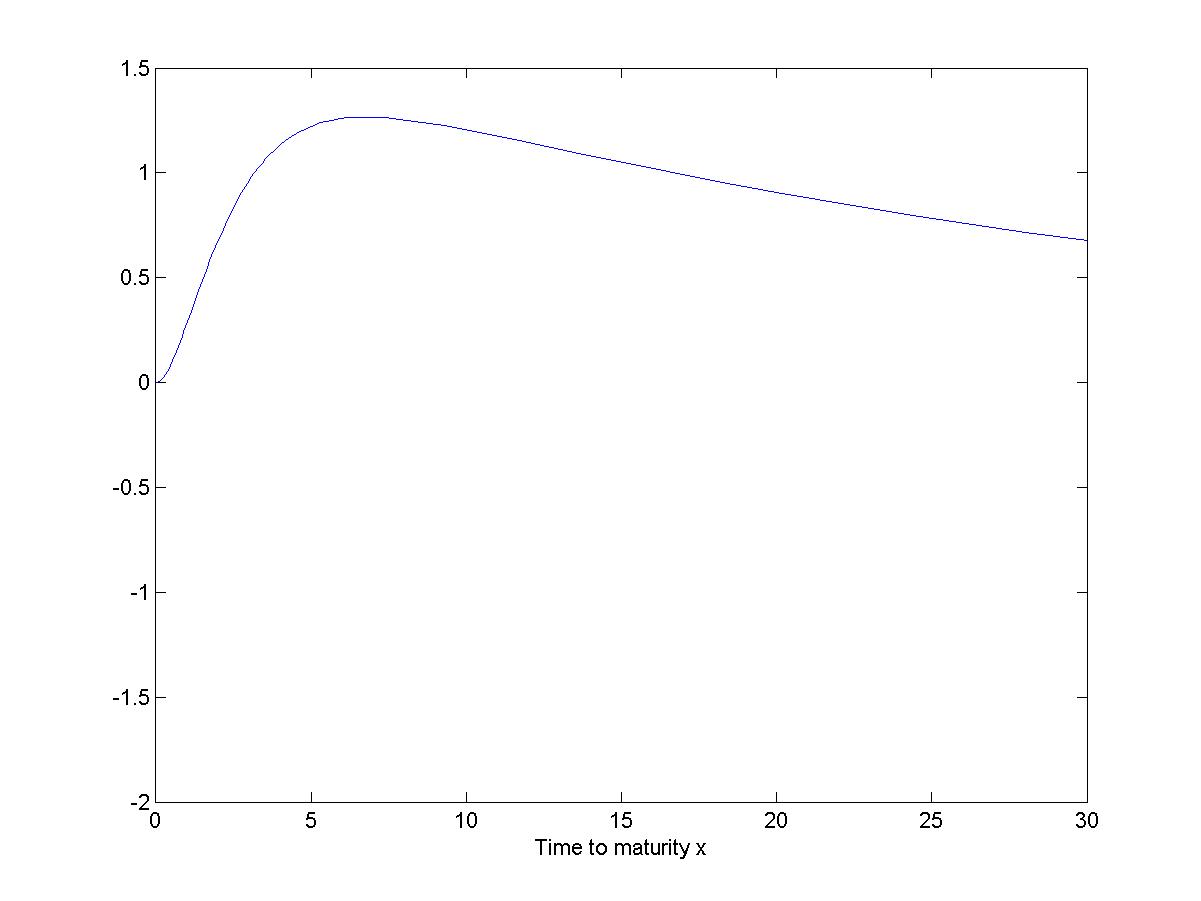}
\caption{graph of $g_2(x)$}
\end{subfigure}
\caption{Graph of the coefficient function $g_i(x)$ with model parameters $\alpha = 0.5, \beta = 0.03, k = 0.1, l = 0.2$. Recall that the time $t$ bond price with maturity $T$ is given by $g_0(T-t) + g_1(T-t) r_t + g_2(T-t) r_t^2$.}
\label{fi:g}
\end{figure}
\begin{figure}
\centering
\begin{subfigure}[b]{0.49\textwidth}
\centering
\includegraphics[width = \textwidth]{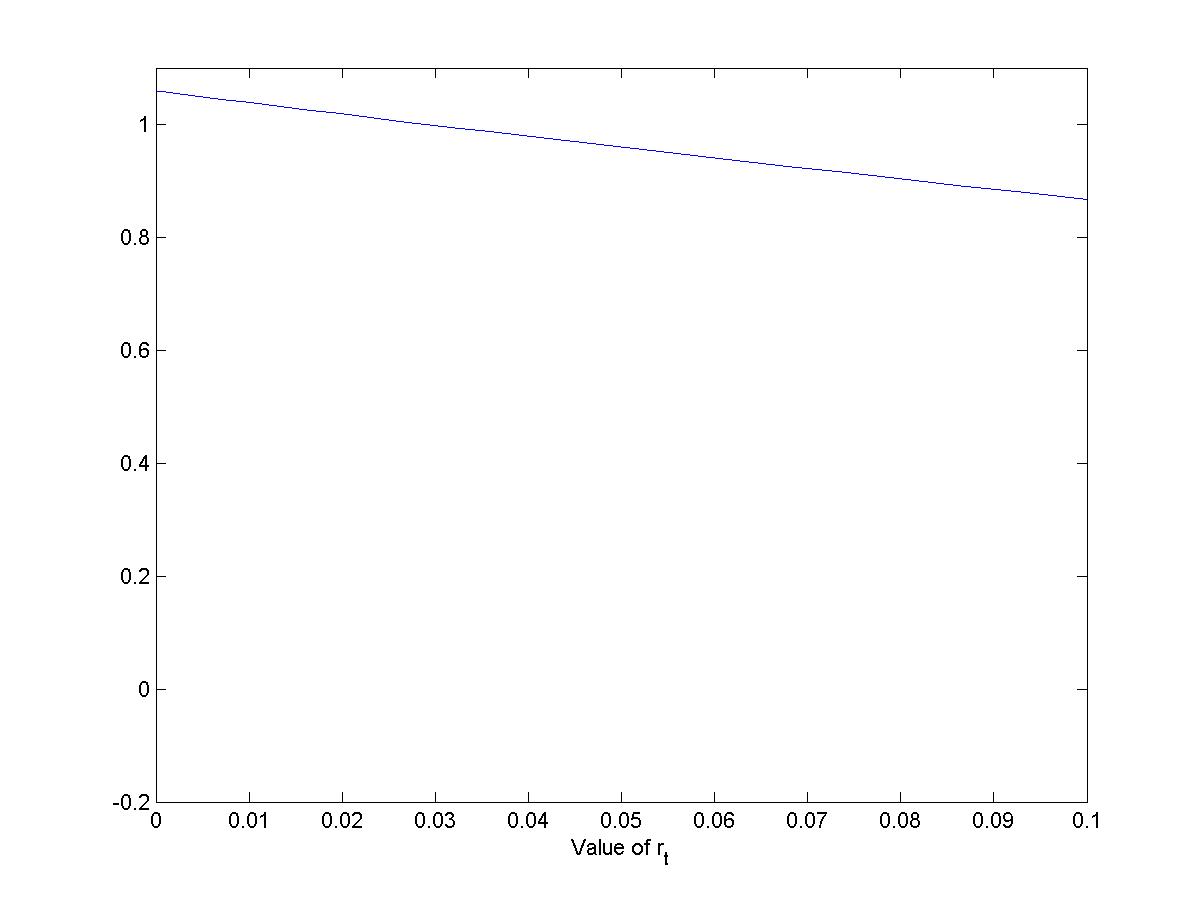}
\caption{graph of $P_0(r)$}
\end{subfigure}
\hfill
\begin{subfigure}[b]{0.49\textwidth}
\centering
\includegraphics[width = \textwidth]{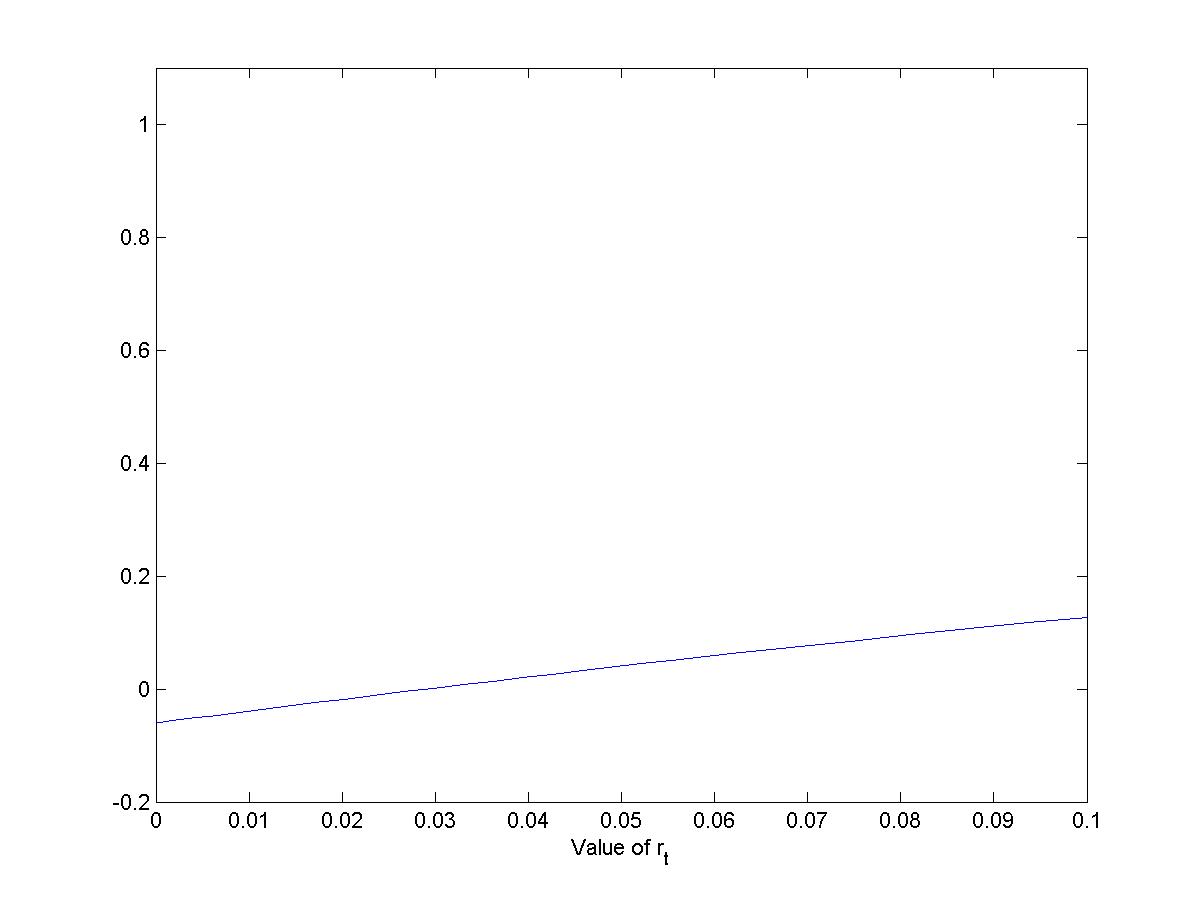}
\caption{graph of $P_1(r)$}
\end{subfigure}
\hfill
\begin{subfigure}[b]{0.5\textwidth}
\centering
\includegraphics[width = \textwidth]{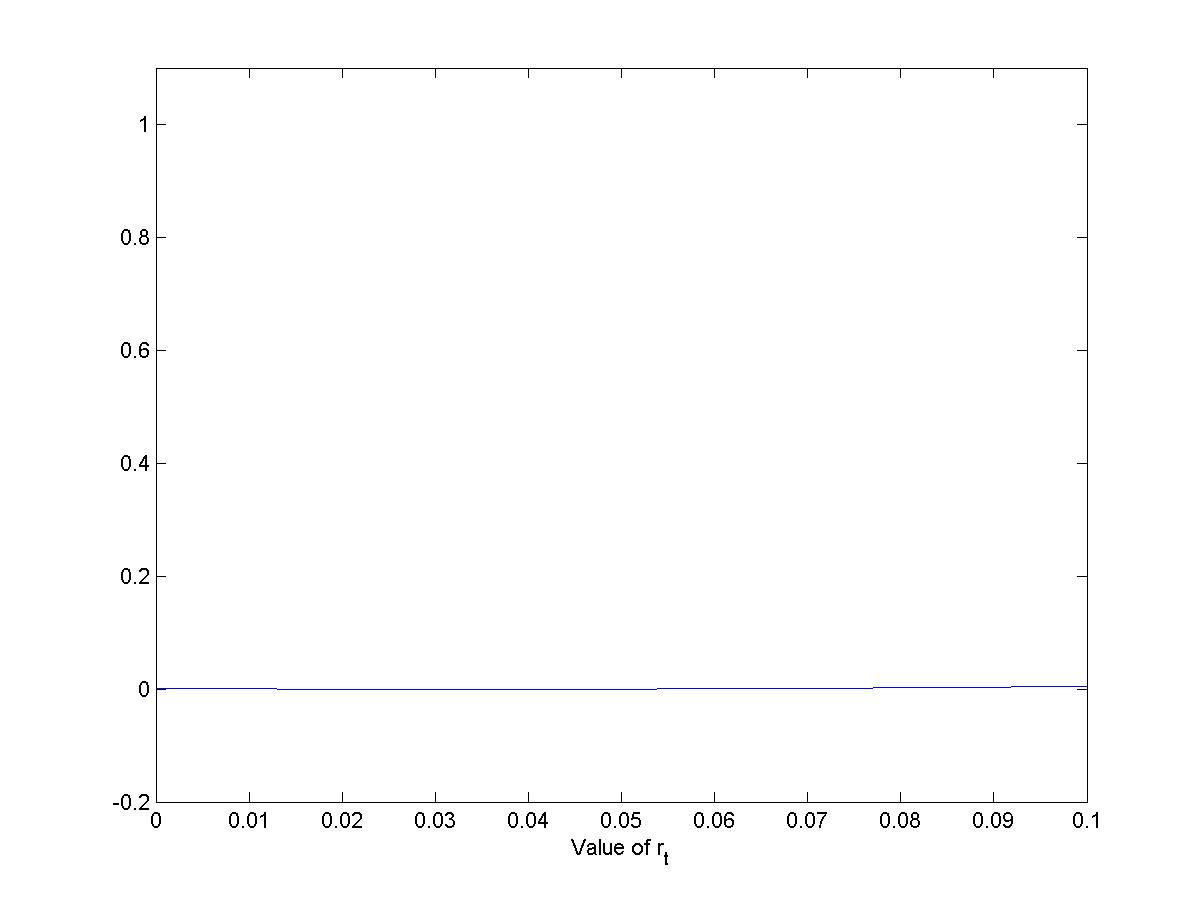}
\caption{graph of $P_2(r)$}
\end{subfigure}
\caption{Graph of the coefficient function $g_i(x)$ with model parameters $\alpha = 0.5, \beta = 0.03, k = 0.1, l = 0.2$. These graph are plotted with domain $r \in (0,0.1)$ chosen in line with the state space of spot rate $r$. Recall that the time $t$ bond price with maturity $T$ is given by $P_0(r_t) e^{-0.0294(T-t)}+ P_1(r_t) e^{-0.5377(T-t)} + P_2(r_t) e^{-1.2329(T-t)}$.}
\label{fi:P}
\end{figure}
\end{ex}

\newpage
The next example is a square-root spot rate model where the factor has the interpretation as the square-root of the spot rate. The dynamics are chosen in such a way that the eigenvalues of the $S$ matrix can be computed explicitly and hence the bond prices can be calculated in closed form.
\begin{ex}[Two-parameter family]
The factor process $Z$ satisfies the following SDE and the spot rate process $r$ is linked to $Z$ through function $R(z) = z^2$.

\begin{align*}
dZ_t &= (Z_t-k)(Z_t+2k+\alpha)(Z_t-2k-\alpha)dt + \sqrt{Z_t^3(2k-Z_t)}dW_t\\
r_t &= Z_t^2
\end{align*}
with parameters $\alpha, k > 0$.

Theorem \ref{th:polyf} says that the factor process $Z_t$ will stay in the open interval $(0,2k)$ as long as:
\begin{align*}
\frac{\alpha(4k+\alpha)}{8k^2} \geq \frac{1}{2}
\end{align*}
The matrix $S$ of this family takes the following form:
\begin{align*}
S = \left(
      \begin{array}{ccc}
        0 & k(2k+\alpha)^2 & 0 \\
        0 & -(2k+\alpha)^2 & 2k(2k+\alpha)^2 \\
        -1 & -k & -2(2k+\alpha)^2 \\
      \end{array}
    \right)
\end{align*}
For ease of notation, we may set $\beta = (2k+\alpha)^2, D=\sqrt{\beta^2-2k^2\beta}$. Then the characteristic polynomial of $S$ is given by:
\begin{align*}
\det(\lambda I -S) = (\lambda + \beta)(\lambda^2+2\beta\lambda+2k^2\beta)
\end{align*}
Hence the eigenvalues of $S$ are $-\beta,-\beta-D,-\beta+D$ and the corresponding eigenvectors will be
\begin{align*}
\left(
  \begin{array}{c}
    -k \\
    1 \\
    0 \\
  \end{array}
\right), \quad \left(
                 \begin{array}{c}
                   -\frac{k\beta}{D+\beta} \\
                   1 \\
                   -\frac{D}{2k\beta} \\
                 \end{array}
               \right), \quad \left(
                                \begin{array}{c}
                                  \frac{k\beta}{D-\beta} \\
                                  1 \\
                                  \frac{D}{2k\beta} \\
                                \end{array}
                              \right)
\end{align*}
Therefore $S$ can be decomposed as $S=PMP^{-1}$, where
\begin{align*}
P &= \left(
       \begin{array}{ccc}
         -k & -\frac{k\beta}{D+\beta} & \frac{k\beta}{D-\beta} \\
         1 & 1 & 1 \\
         0 & -\frac{D}{2k\beta} & \frac{D}{2k\beta} \\
       \end{array}
     \right) \quad M = \left(
                         \begin{array}{ccc}
                           -\beta & 0 & 0 \\
                           0 & -\beta-D & 0 \\
                           0 & 0 & -\beta+D \\
                         \end{array}
                       \right)
     \\
P^{-1} &= D^{-1}(\frac{1}{2k^2}-\frac{1}{\beta})^{-1}\left(
                                                       \begin{array}{ccc}
                                                         \frac{D}{k\beta} & \frac{D}{2k^2} & \frac{D}{k} \\
                                                         -\frac{D}{2k\beta} & -\frac{D}{2\beta} & \frac{kD}{D-\beta} \\
                                                         -\frac{D}{2k\beta} & -\frac{D}{2\beta} & -\frac{kD}{D+\beta} \\
                                                       \end{array}
                                                     \right)
\end{align*}
Recall that the coefficient functions $(g_i(x))_i$ satisfy a matrix ODE $\dot{G}(x) = SG(x)$. Solving the ODE gives:
\begin{align*}
\left(
  \begin{array}{c}
    g_0(x) \\
    g_1(x) \\
    g_2(x) \\
  \end{array}
\right) = e^{Sx} \left(
                   \begin{array}{c}
                     1 \\
                     0 \\
                     0 \\
                   \end{array}
                 \right)
\end{align*}
The coefficient functions $(g_i(x))_i$ take the following explicit form:
\begin{align*}
\left(
  \begin{array}{c}
    g_0(x) \\
    g_1(x) \\
    g_2(x) \\
  \end{array}
\right) = \frac{k}{2k^2-\beta} \left(
                   \begin{array}{c}
                     2ke^{-\beta x}-\frac{k\beta}{D+\beta}e^{(-\beta-D)x}+\frac{k\beta}{D-\beta}e^{(-\beta+D)x} \\
                     -2e^{-\beta x}+e^{(-\beta-D)x}+e^{(-\beta+D)x} \\
                     -\frac{D}{2k\beta}e^{(-\beta-D)x}+\frac{D}{2k\beta}e^{(-\beta+D)x} \\
                   \end{array}
                 \right)
\end{align*}
Especially, given initial spot rate $r_0$, the time 0 bond price and yield with time to maturity $x$ can be expressed as:
\begin{align*}
P(x,r_0) &= g_0(x) + g_1(x)\sqrt{r_0} + g_2(x) r_0\\
y(x,r_0) &= -\frac{\log{P(x,r_0)}}{x}
\end{align*}

For calibration, we again take the one month yield as an approximation to the initial spot rate and calculate the complete theoretical yield curve of this model. We then adjust the values of the two parameters $\alpha,k$ and try to fit the remaining 4,300 observations. By performing 2,000 random searches in the range $\alpha \in (0,1)$ and $k \in (0,1)$, we get the best result is $\alpha = 0.172, k = 0.206$ with error $E_1 = 0.0902$. Hence the average difference is
$$
\text{average difference \%} = \sqrt{0.0902/4300} \times 100\% = 0.46\%
$$
Notice that this choice of parameter satisfies the condition in theorem \ref{th:polyf}. The matrix $S$ is given by:
\begin{align*}
S = \left(
      \begin{array}{ccc}
        0 & 0.0703 & 0 \\
        0 & -0.3411 & 0.1405 \\
        -1 & -0.2060 & -0.6821 \\
      \end{array}
    \right)
\end{align*}
With the above parameter values, the simulated the spot rate process $r_t$ and yield curve with $r_0 = 10^{-4}$ are shown in Figures \ref{fi:spot1} and \ref{fi:yield1}:
\begin{figure}
\centering
\includegraphics[scale=0.3]{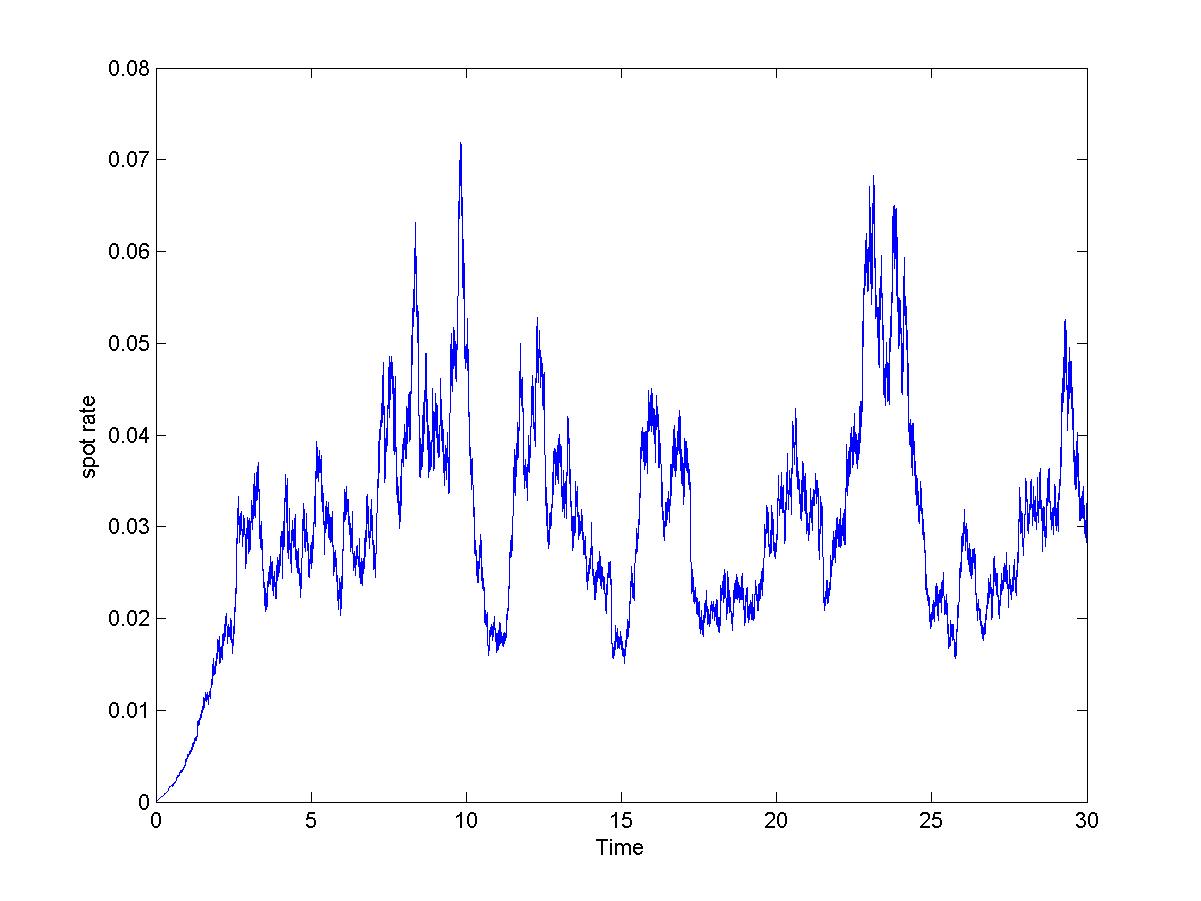}
\caption{Typical sample path of the spot rate process $r$. Where $r_t = \sqrt{Z_t}$ and
$
dZ_t = (Z_t-k)(Z_t+2k+\alpha)(Z_t-2k-\alpha)dt + \sqrt{Z_t^3(2k-Z_t)}dW_t
$
with model parameter $\alpha = 0.172, k =0.206$. The initial spot rate is set to $r_0 = 0.01\% $ in line with current situation.}
\label{fi:spot1}
\end{figure}
\begin{figure}
\centering
\includegraphics[scale = 0.25]{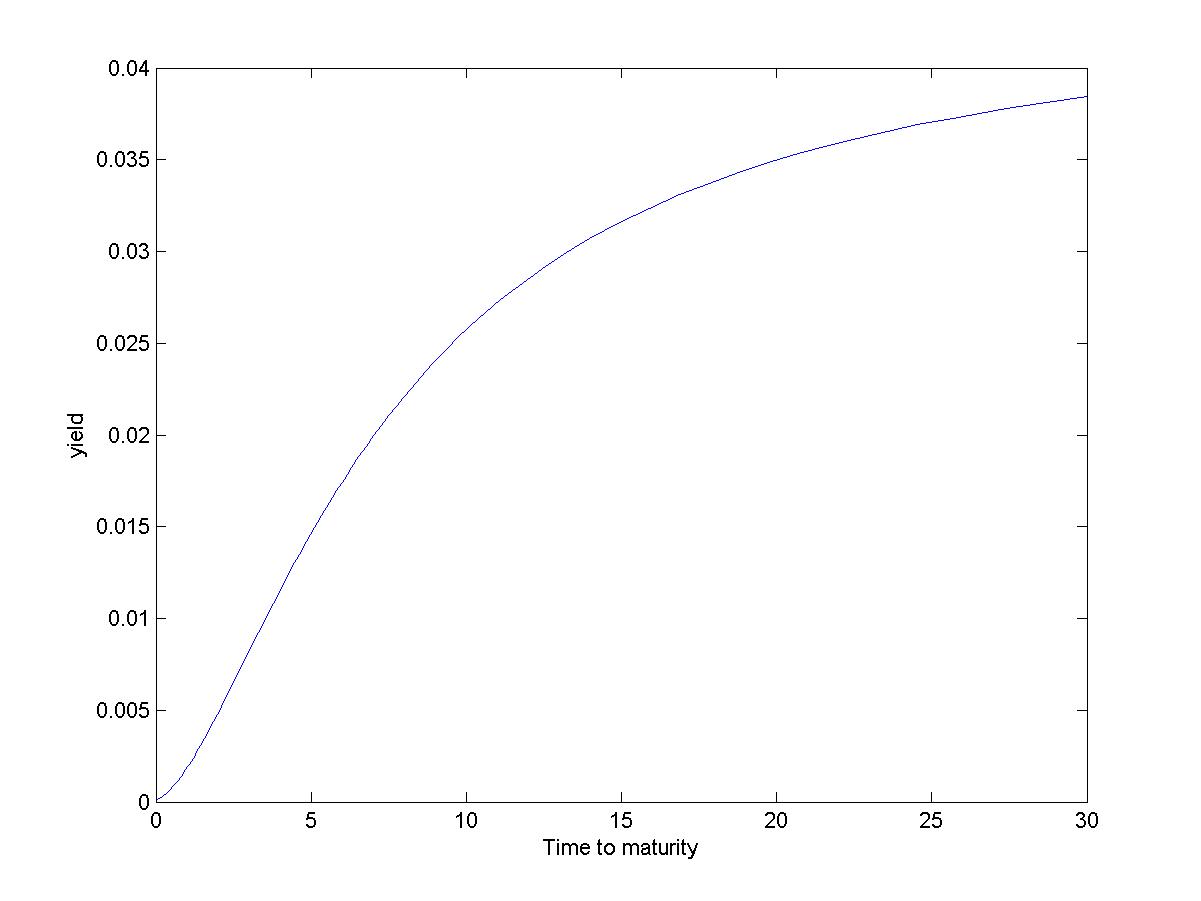}
\caption{Yield curve from current date to 30 years generate by model parameter $\alpha = 0.172, k = 0.206$. Initial spot rate is set to $r_0 = 0.01\%$.}
\label{fi:yield1}
\end{figure}

By changing the initial spot rate $r_0$, we can also get different shapes of yield curve as shown in Figure \ref{fi:yield3}:
\begin{figure}
\centering
\includegraphics[scale = 0.25]{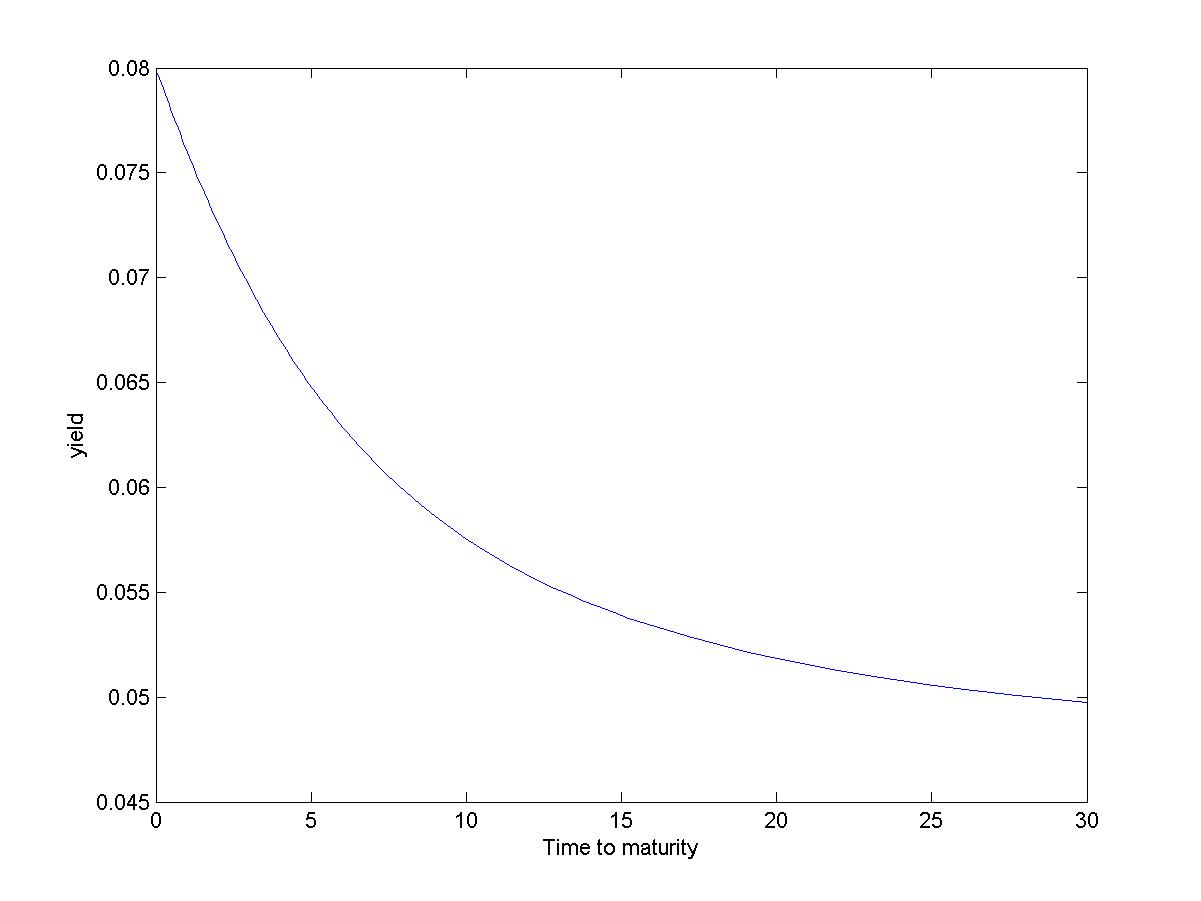}
\caption{Yield curve from current date to 30 years generate by model parameter $\alpha = 0.172, k = 0.206$. Initial spot rate is set to $r_0 = 8\%$.}
\label{fi:yield3}
\end{figure}

The corresponding eigenvalues are (numerically): $-0.0455, -0.6366, -0.3411$\\
The graph of $g_i,P_i$ function are shown in Figures \ref{fi:g1} and \ref{fi:p1}:
\begin{figure}
\centering
\begin{subfigure}[b]{0.49\textwidth}
\centering
\includegraphics[width = \textwidth]{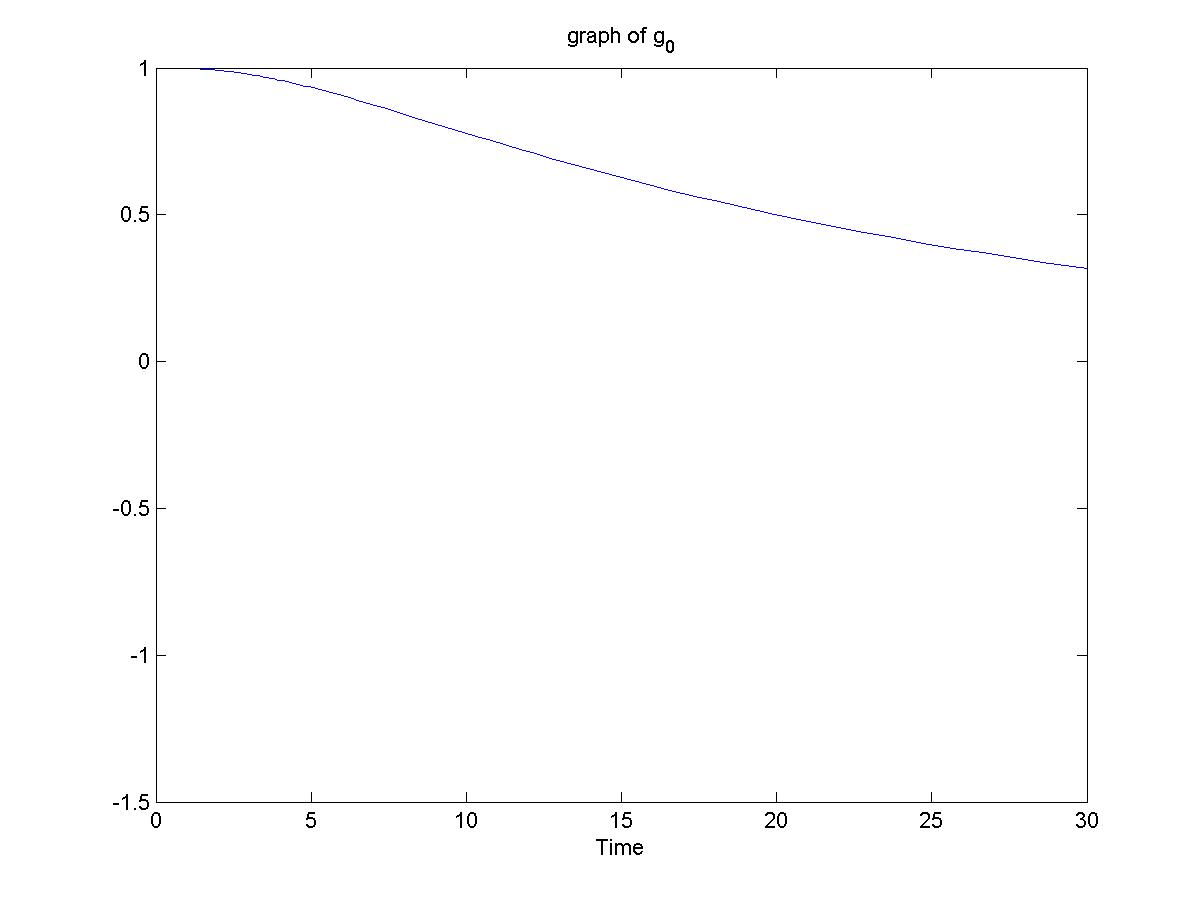}
\caption{graph of $g_0(x)$}
\end{subfigure}
\hfill
\begin{subfigure}[b]{0.49\textwidth}
\centering
\includegraphics[width = \textwidth]{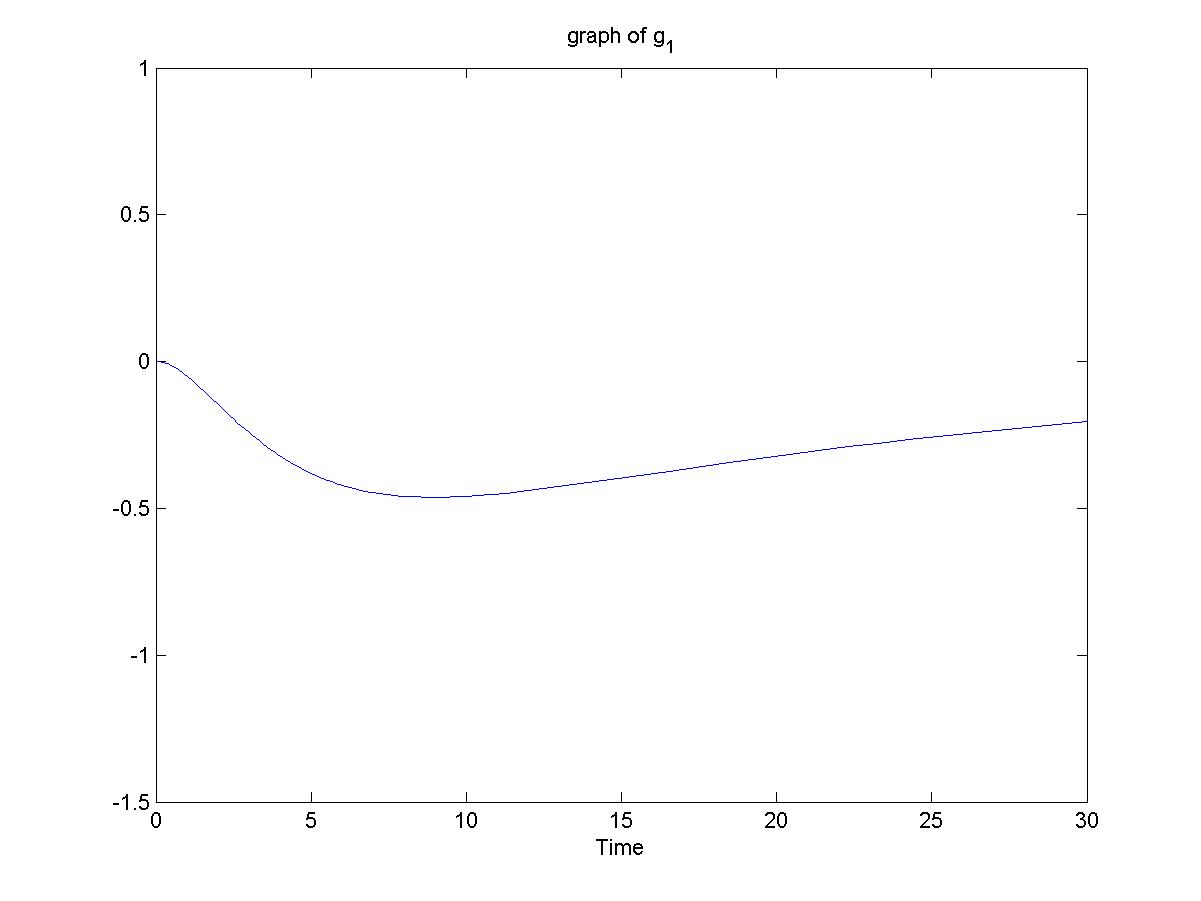}
\caption{graph of $g_1(x)$}
\end{subfigure}
\hfill
\begin{subfigure}[b]{0.5\textwidth}
\centering
\includegraphics[width = \textwidth]{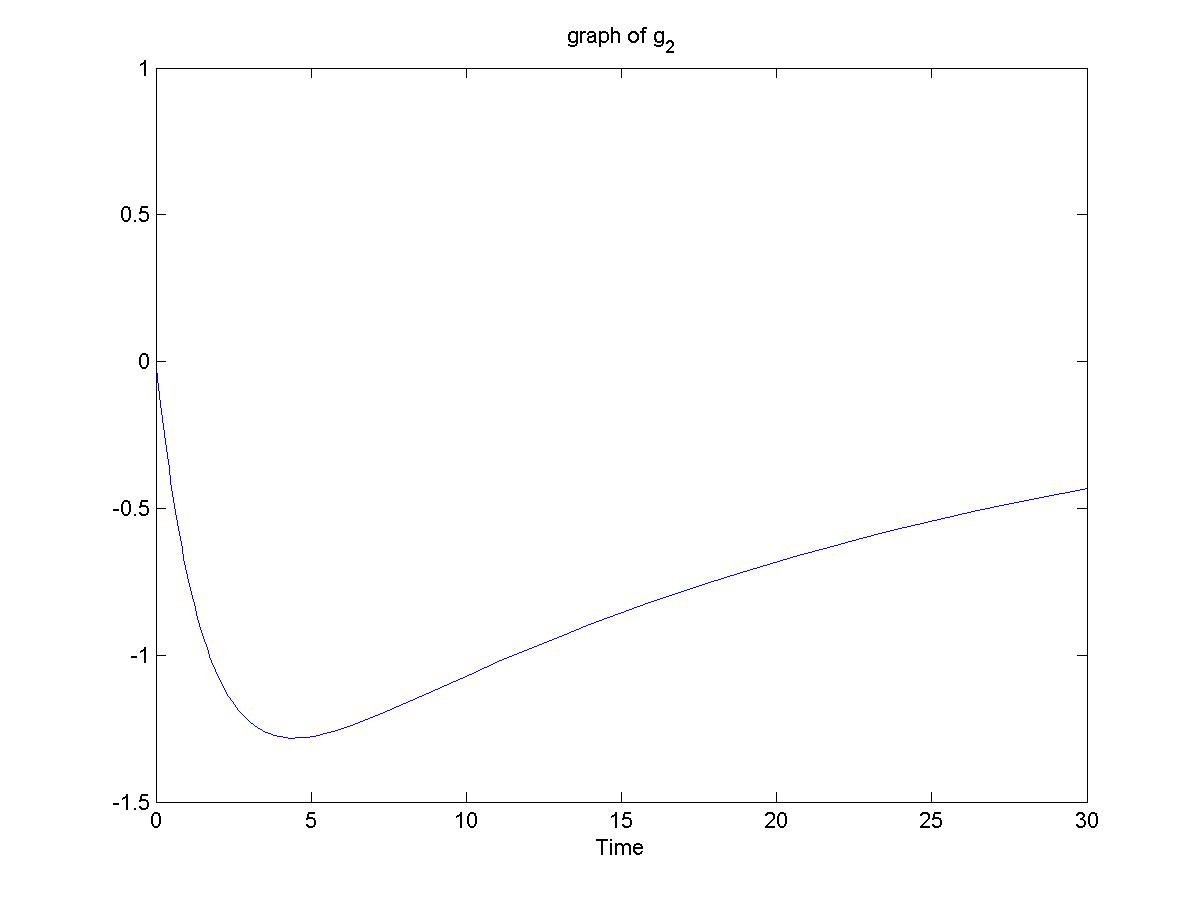}
\caption{graph of $g_2(x)$}
\end{subfigure}
\caption{Graph of the coefficient functions $g_0(x), g_1(x), g_2(x)$ on the same scale of time range $[0,30]$ with model parameter $\alpha = 0.172, k =0.206$. The time $t$ price of bond maturing at time $T$ in this model is given by $g_0(T-t) + g_1(T-t) \sqrt{r_t} + g_2(T-t) r_t$.}
\label{fi:g1}
\end{figure}

\begin{figure}
\centering
\begin{subfigure}[b]{0.49\textwidth}
\centering
\includegraphics[width = \textwidth]{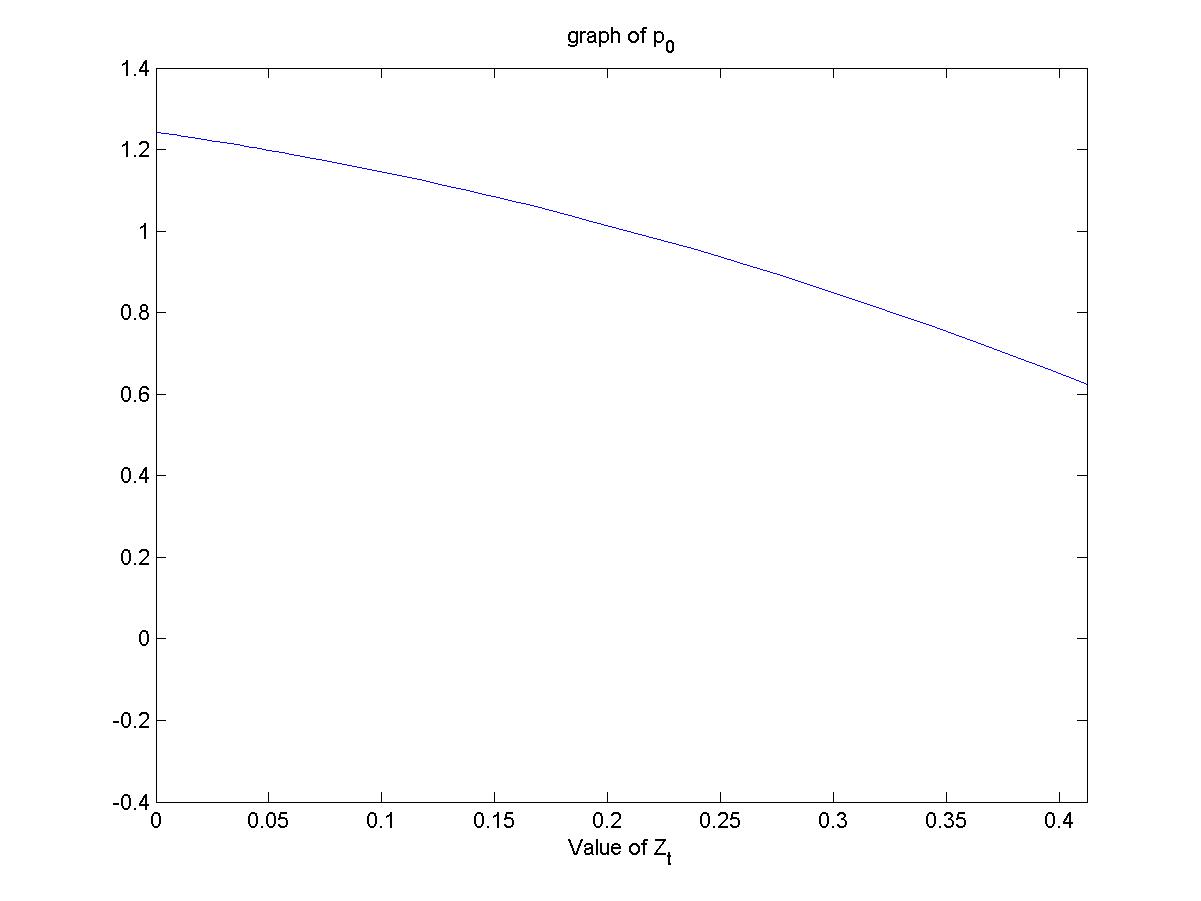}
\caption{graph of $P_0(z)$}
\end{subfigure}
\hfill
\begin{subfigure}[b]{0.49\textwidth}
\centering
\includegraphics[width = \textwidth]{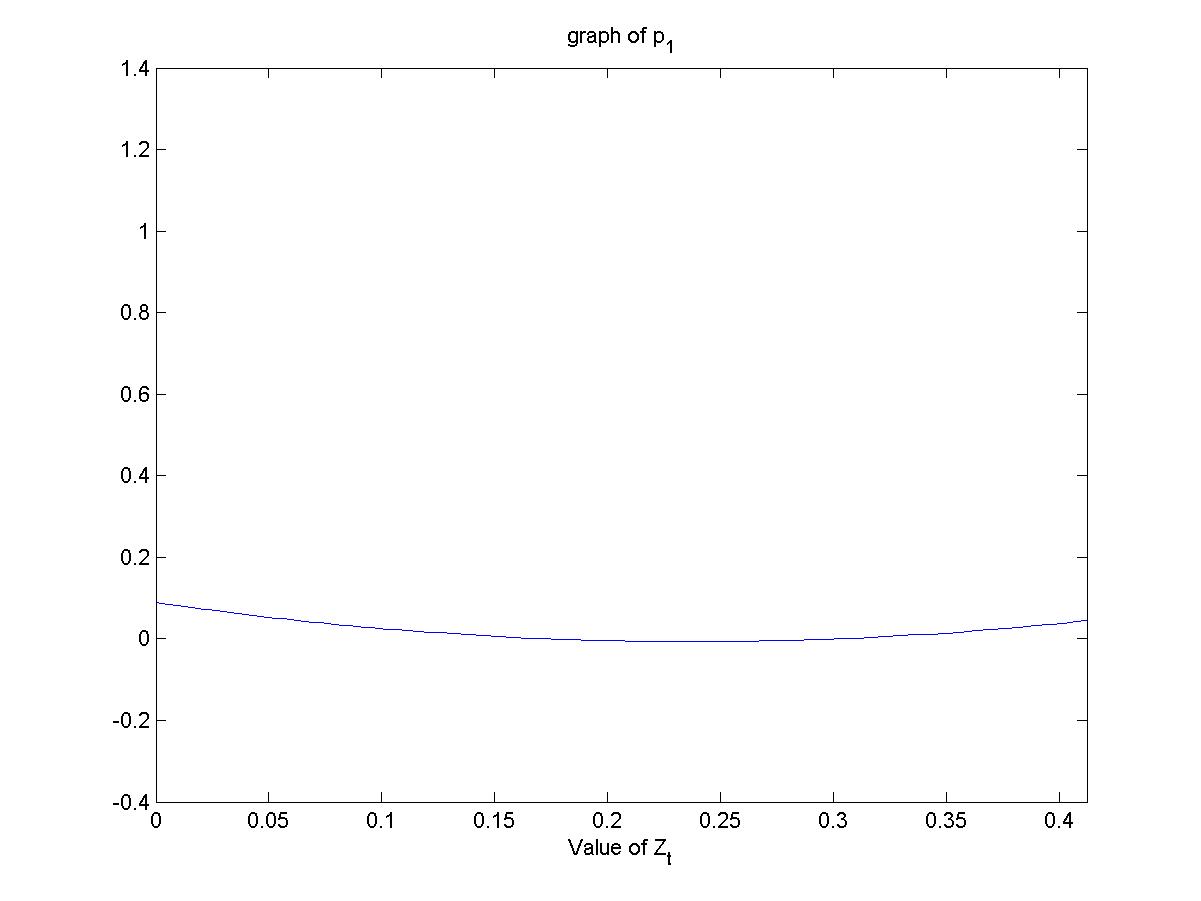}
\caption{graph of $P_1(z)$}
\end{subfigure}
\hfill
\begin{subfigure}[b]{0.5\textwidth}
\centering
\includegraphics[width = \textwidth]{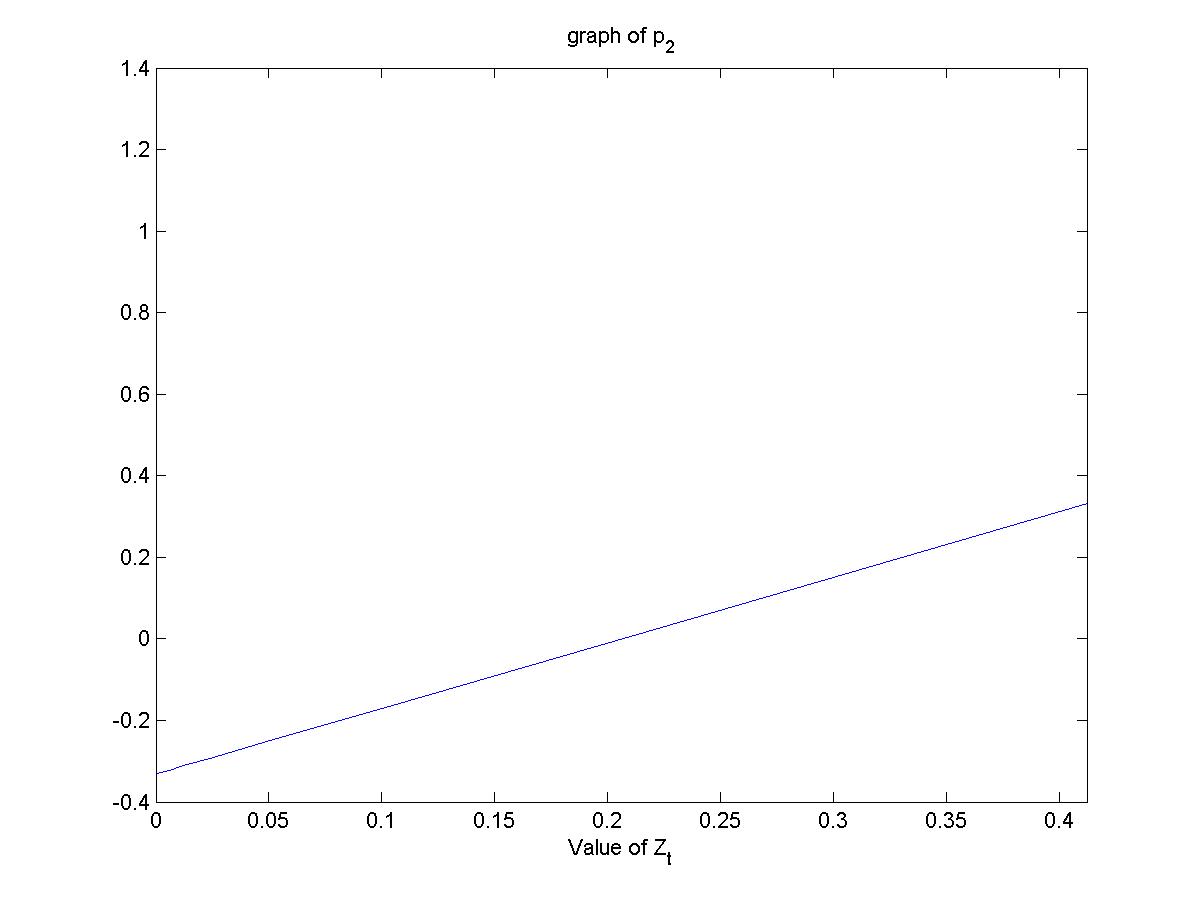}
\caption{graph of $P_2(z)$}
\end{subfigure}
\caption{Graph of the functions $P_0(z), P_1(z), P_2(z)$ with model parameters $\alpha = 0.172, k = 0.206$. Since the state space of the factor process $Z$ is $(0,2k)$, we choose to plot the graph on the same scale with $z \in (0,0.412)$. The time $t$ price of bond maturing at time $T$ in this model is given by $P_0(Z_t) e^{-0.0455(T-t)}+ P_1(Z_t) e^{-0.6366(T-t)} + P_2(Z_t) e^{-0.3411(T-t)}$ and the spot rate is given by $r_t = Z_t^2$.}
\label{fi:p1}
\end{figure}
\end{ex}

\newpage

\begin{remark}
Notice that all eigenvalues of the S matrix are real and negative in both examples above, hence the bond price can be viewed as a linear combination of bond prices with fixed positive interest rate given by $r = -\lambda_i$.
\end{remark}

Finally in this section, we fit the data in table \ref{tab:yahoo} with the famous CIR \cite{CIR} model. First recall that CIR model is a three-parameter spot rate model where the spot rate $r$ satisfies:
\begin{align*}
dr_t = a(b - r_t) dt + \sigma \sqrt{r_t} dW_t
\end{align*}
The bond price can be solved explicitly as
\begin{align*}
P_t(T) = \exp{(A(T-t) + B(T-t) r_t)}
\end{align*}
where
\begin{align*}
B(x) &= \frac{2(1-e^{hx})}{2h + (a+h)(e^{hx - 1})}\\
A(x) &= \frac{2ab}{\sigma^2} \left( \log \left(\frac{2he^{\frac{(a+h)x}{2}}}{2h+(a+h)(e^{hx}-1)} \right)\right)\\
h &= \sqrt{a^2 + 2\sigma^2}
\end{align*}
By performing 5,000 random searches over the region $a \in (0,3),b \in (0,0.2), \sigma^2 \in (0,1)$, the best result is $a = 0.6443, b = 0.0254, \sigma^2 = 0.0251$ with error $E = 0.4957$. Hence the average difference is
$$
\text{average difference \%} = \sqrt{0.4957/4300} \times 100\% = 1.07\%
$$
Let's summarize three parametric family of tractable interest rate models in the table below:
\begin{table}[h]
\centering
\begin{tabular}{l|l|l|l|l|l}
Model & Type & Total degree & Nr. of parameters & Sum of squares & Average error \% \\
\hline
Ex1& Polynomial & 2& 4 &  0.3246 & 0.87 \\
Ex2& Polynomial & 2& 2 &  0.0902 & 0.46 \\
CIR& Exp. polynomial & N/A& 3 & 0.4957 & 1.07
\end{tabular}
\caption{A comparison of two quadratic models discussed in this section with CIR model. All three models are calibrate to match US Treasury bond yield from 2006 Feb. $10^{\text{th}}$ to 2014 May $9^{\text{th}}$ sampled weekly with a total of $430 \times 11$ data. It is not surprising that EX1 outperforms CIR because the former has one more parameter than the latter and hence may fit the data better. Interestingly in this case, EX2 also outperforms CIR with only two parameters. Actually EX2 has the best performance with the least number of parameters in this case.}
\label{tab:comp}
\end{table}

\section{Extensions} \label{se:extensions}
In this section, we will extend theorem \ref{th:main} in two different ways: namely allowing time dependency and allowing a multi-dimensional factor process.
\subsection{Hull-White extension}
As usual, by incorporating time-dependent parameters, we can hope to have a better model calibration. We introduce time dependency both in the dynamics of the factor process $(Z_t)_{t \geq 0}$ and the coefficient functions $g_k$. As one may expect, we will establish a similar sufficient and necessary condition in this case.

To be clear, we now consider a factor process $(Z_t)_{t \geq 0}$ be a non-explosive solution to the following time-inhomogeneous SDE
$$
dZ_t = b(t,Z_t)dt + \sigma(t,Z_t)dW_t
$$
The spot rate $r_t$ is modelled as $r_t = R(t,Z_t)$ and the bond price and discounted bond price are defined by
\begin{align*}
P_t(T) &= \sum_{k=0}^n g_k(t,T)Z_t^k \\
\tilde{P_t}(T) &= e^{-\int_0^t R(s,Z_s)ds}P_t(T)
\end{align*}
where $g_k: \Delta \to \RR$ are smooth deterministic functions satisfying the boundary conditions:
\begin{align*}
g_0(T,T) &= 1 \\
g_k(T,T) &= 0 \quad \text{for all} \quad 1 \leq k \leq n
\end{align*}
where $\Delta = \{ (t,T): 0 \le t \le T \}$.

By adding the $t$ component, the consistent PDE \eqref{eq:PDE} becomes:
\begin{equation} \label{eq:cons}
\sum_{k=0}^n \frac{\partial g_k}{\partial t}(t,T) z^k = \sum_{k=0}^n g_k(t,T) A_k(t,z) \quad \forall (t,T,z)
\end{equation}
Where $A_k(t,z)$ are defined as:
\begin{align*}
A_k(t,z) &:= R(t,z)z^k - kb(t,z)z^{k-1} - \frac{k(k-1)}{2}a(t,z)z^{k-2} \\
a(t,z)   &:= \sigma^2(t,z)
\end{align*}

\begin{theorem}Suppose that $n\ge2$ and that the functions $g_k(t, \cdot)$ are linearly independent for all $t \geq 0$. Then we must have $R(t,z) = \sum_{i=0}^2 R_i(t)z^i$, $b(t,z) = \sum_{i=0}^3 b_i(t) z^i$, $a(t,z) = \sum_{i=0}^4 a_i(t) z^i$ where the coefficients satisfy
\begin{align*}
\frac{n(n-1)}{2}a_4(t) + nb_3(t) - R_2(t) &= 0 \\
\frac{(n-1)(n-2)}{2}a_4(t) + (n-1)b_3(t) - R_2(t) &= 0\\
\frac{n(n-1)}{2}a_3(t) + nb_2(t) - R_1(t) &= 0
\end{align*}
and the coefficient functions $g_k$ are determined by the unique solution to the ODE
\begin{align*}
\frac{\partial}{\partial t} G(t,T) &= S(t) G(t,T)\\
G(T,T) &= (1,0,\ldots,0)^\top
\end{align*}
and the $(n+1) \times (n+1)$ matrix $S(t)$ is defined by
\begin{align*}
S_{j+k,j}(t) &= R_k(t) - jb_{k+1}(t) - \frac{j(j-1)}{2}a_{k+2}(t)
\end{align*}
\end{theorem}

\begin{proof}
Fix any $t$, choose $t < T_0(t) < \ldots < T_n(t)$ such that we can rewrite the consistency condition \eqref{eq:cons} as follows:
\begin{align*}
\left(
  \begin{array}{ccc}
    g_0(t,T_0(t)) & \cdots & g_n(t,T_0(t)) \\
    \vdots & \ddots & \vdots \\
    g_0(t,T_n(t)) & \cdots & g_n(t,T_n(t)) \\
  \end{array}
\right) \left(
          \begin{array}{c}
            A_0(t,z) \\
            \vdots \\
            A_n(t,z) \\
          \end{array}
        \right)
&= \left(
     \begin{array}{c}
       \sum_{i=0}^n \frac{\partial g_i}{\partial t}(t,T_0(t))z^i \\
       \vdots \\
       \sum_{i=0}^n \frac{\partial g_i}{\partial t}(t,T_0(t))z^i \\
     \end{array}
   \right)
\end{align*}
Which is of the form
\begin{align*}
M(t)x(t,z) &= c(t,z)
\end{align*}
Since the functions $g_i(t,\cdot)$ are linearly independent, we can choose $T_i(t)$ such that the matrix $M(t)$ is invertible. Hence the solution is given by
\begin{align*}
x(t,z) &= M^{-1}(t)c(t,z)
\end{align*}
But for fixed $t$, $c(t,z)$ consists of linear combinations of $z^0,\ldots,z^n$, we deduce that $x(t,z)$ must be a linear combinations of $z^0,\ldots,z^n$. Hence each $A_i(t,z)$ must be the form of $\sum_{i=0}^n D_i(t)z^i$. i.e. only the coefficient of $z^i$ depends on $t$. The rest of the proof goes exactly the same as the time independent case.
\end{proof}

\subsection{Multi-dimensional factor process}
In this subsection, we will extend the polynomial model framework by allowing both factor process $(Z_t)_{t \geq 0}$ and the background Brownian motion $(W_t)_{t \geq 0}$ to be multi-dimensional. To be more specific, let $(W_t)_{t \geq 0}$ be a  $D$-dimensional Brownian motion. Let $(Z_t)_{t \geq 0}$ be the factor process taking values in $I \subseteq \RR^d$, assuming to be the (non-explosive) solution of the SDE:
$$
dZ_t =  b(Z_t)dt + \sigma(Z_t)dW_t
$$
for some continuous deterministic functions $b : \RR^d \to \RR^d$ and $\sigma : \RR^d \to \RR^{d \times D}$. We define the diffusion function $a = \sigma \sigma^\top$, and note that the only role played by the parameter $D$ is as the upper bound on the rank of the matrix $a(z)$.

For $k = (k_1, \ldots , k_d) \in \mathbb{Z}^d_+$ and $z = (z_1, \ldots, z_d) \in \RR^d$, we define the monomial $z^k$ as follows:
$$
z^k := z_1^{k_1}\cdots z_d^{k_d}
$$
We define the total degree of $k$ to be $|k| = k_1 + \ldots +k_d$, and set $K_n = \{ k \in \mathbb{Z}^d_+ : |k| \le n \}$. With the notation defined above, we let the bond price to be:
$$
P_t(T) = \sum_{k \in K_n} g_k(T-t) Z_t^k
$$
where the functions $g_k$ satisfy the boundary conditions
\begin{align*}
g_k(0) = 1 \quad \text{if } |k| = 0 \\
g_k(0) = 0 \quad \text{otherwise}
\end{align*}
The spot rate is modelled similarly as $r_t = R(Z_t)$ for some deterministic function $R : \RR^d \to \RR$.

The no arbitrage condition \eqref{eq:PDE} turns out to be the condition:
\begin{equation} \label{eq:md}
\sum_{k \in K_n} \dot{g}_k(x) z^k = \sum_{k \in K_n} g_k(x) A_k(z)
\end{equation}
holds for any $x \ge 0$ and $z \in I$, where the functions $A_k$ are defined as
\begin{displaymath}
A_k(z) = \sum_{i=1}^d b_i(z) \frac{\partial (z^k)}{\partial z_i} + \frac{1}{2} \sum_{i,j = 1}^d a_{ij}(z) \frac{\partial^2 (z^k)}{\partial z_i \partial z_j} - R(z)z^k
\end{displaymath}

Finally we define the notation
$$
F_n = \left\{ f(z) : \sum_{k \in K_n} f_k z^k, \quad f_k \in \RR \right\}
$$
to be the family of polynomials in $d$ variables of total degree less or equal to $n$.

\begin{theorem}
Suppose $n \ge 2$ and that the functions $(g_k)_k$ are linearly independent. Then we must have $R \in F_2$, $b_i \in F_3 \quad 1 \le i \le d$ and $a_{ij} \in F_4 \quad 1 \le i,j \le d$. Furthermore, the coefficients are constrained in such a way that $A_k \in F_n$ for all $k$ such that $|k| \in \{n-1,n\}$.
\end{theorem}
\begin{proof}
First we show that the functions $A_k \in F_n$ are polynomials for all $k \in K_n$. Let $N = |K_n|$ be the cardinality of set $K_n$. Since the functions $(g_k)_k$ are linearly independent, we can find $N$ distinct points $x_1, \ldots, x_N$ independent of $z$ such that the matrix with $i$-th column formed by vector $(g_k(x_i) , k \in K_n)$ is non-singular. Now fix any $z$, we can rewrite the no-arbitrage condition \eqref{eq:md} as a set of $N$ simultaneous linear equations with $N$ unknowns $A_k(z)$. Therefore the solution exists and is unique and can be written as linear combinations of the monomials $z^k$, hence all of the $A_k(z)$ are polynomials in $d$ variables of total degree less or equal to $n$.

For ease of notation, we introduce the following definition:
\begin{align*}
(a)_i & := (0, \ldots, 0, a, 0, \ldots , 0) && \text{where $a$ is the $i$-th component.}\\
(a,b)_{i,j} & := (0, \ldots, a, \ldots, b, \ldots, 0) &&\text{where $a$ is the $i$-th component and $b$ is the $j$-th component.}\\
\end{align*}
Since we must have $A_k(z) \in F_n$ for all $k \in K_n$, we can conclude for any $1\leq i,j \leq d$
\begin{align*}
A_0(z) &= -R(z) && \in F_n \\
A_{(1)_i}(z) &= b_i(z) - z_i R(z) && \in F_n \\
A_{(1,1)_{i,j}}(z) &= b_i(z) z_j + b_j(z) z_i + a_{ij}(z) - z_i z_j R(z) && \in F_n
\end{align*}
Therefore we may conclude immediately that the functions $R, b_i, a_{ij}$ are polynomials. On the other hand
\begin{align*}
A_{(n)_i}(z) = n z_i^{n-1}b_i(z) + \frac{n(n-1)}{2} z_i^{n-2}a_{ii}(z) - z_i^n R(z) \in F_n
\end{align*}
by cancelling the $z_i^{n-2}$ factor, we may deduce that:
\begin{equation} \label{eq:md1}
nz_ib_i(z) + \frac{n(n-1)}{2} a_{ii}(z) - z_i^2 R(z) \in F_2
\end{equation}
Similarly by considering $A_{(n-1)_i}, A_{(n-2)_i}, A_{(n-1,1)_{i,j}}$, we get
\begin{equation} \label{eq:md2}
(n-1)z_ib_i(z) + \frac{(n-2)(n-1)}{2} a_{ii}(z) - z_i^2 R(z) \in F_3
\end{equation}
\begin{equation} \label{eq:md3}
(n-2)z_ib_i(z) + \frac{(n-2)(n-3)}{2} a_{ii}(z) - z_i^2 R(z) \in F_4
\end{equation}
\begin{equation}
(n-1)z_i z_j b_i(z) + z_i^2 b_j(z) +  \frac{(n-2)(n-1)}{2} z_j a_{ii}(z) + (n-1)z_i a_{ij}(z) - z_i^2 z_j R(z) \in F_3
\end{equation}
Subtracting \eqref{eq:md1} from \eqref{eq:md2} and subtracting \eqref{eq:md2} from \eqref{eq:md3} gives
\begin{align*}
z_ib_i(z) + (n-1) a_{ii}(z)  \in F_3 \\
z_ib_i(z) + (n-2) a_{ii}(z)  \in F_4 \\
\end{align*}
Hence we get the required degree constraint on functions $R, b, a$. For the remaining part of this theorem, observe that given the degree constraint, the functions $A_k$ will automatically $\in F_n$ as long as $|k| \leq n-2$.
\end{proof}
\begin{remark}
We note that the case when $n=1$ essentially is covered in the paper of Siegel.\cite{S}
\end{remark}
\section{Acknowledgement}
The authors acknowledge the financial support of the Man Group studentship and the Cambridge Endowment for Research in Finance.

\end{document}